\documentclass[a4paper,12pt]{article}
\usepackage{graphicx,amssymb}
\renewcommand{\figurename}{Fig.}

\title{Isovector response and energy-weighted sums in hot nuclei}

\author{V. M. Kolomietz, S. V. Lukyanov, O. I. Davidovskaya \\
{\it Institute for Nuclear Research, 03680 Kyiv, Ukraine}}

\date{\empty}

\begin{document}
\maketitle

\begin{abstract}
We investigate the collective response function and the energy-weighted sums
(EWS) $m_{k}$ for isovector mode in hot nuclei. The approach is based on the
collisional kinetic theory and takes into consideration the temperature and
the relaxation effects. We have evaluated the temperature dependence of the
adiabatic, $E_{1}=\sqrt{m_{1}/m_{-1}}$, and scaling, $E_{3}=\sqrt{m_{3}/m_{1}}$,
energy centroids of the isovector giant dipole resonances (IVGDR). The
centroid energy $E_{3}$ is significantly influenced by the Fermi surface
distortion effects and, in contrast to the isoscalar mode, shows much weaker
variation with temperature. Taking into account a connection between the
isovector sound mode and the corresponding surface vibrations we have
established the $A$-dependence of the IVGDR centroid energy which is in a good
agreement with experimental data. We have shown that the enhancement factor
for the "model independent" sum $m_{1}$ is only slightly sensitive to the
temperature change.
\bigskip


\end{abstract}

\section{Introduction}

Many features of nuclei are sensitive to nuclear heating. The nuclear
heating influences strongly the particle distribution near the Fermi surface
and reduces the Fermi-surface distortion effects on the nuclear collective
dynamics \cite{kosh05}. Moreover, the heating of the nucleus provides the
transition from the rare- to frequent interparticle collision regime. One
can expect that the zero-sound excitation modes which exist in cold nuclei
will be transformed to the first-sound ones in hot nuclei. Knowledge of the
nuclear collective dynamics in hot nuclei allows one to understand a number
of interesting phenomena, e.g., the temperature dependence of the basic
characteristics of isovector giant dipole resonance (IVGDR). The existence
of the IVGDR in the heated nuclei built on the excited states was established
a long time ago \cite{snov86}-\cite{baum98}.
A systematic experimental and theoretical study of the IVGDR in hot nuclei
provides a significant information on the isovector collective motion at
non-zero temperatures \cite{sabl06,schitho07}.

A good first orientation in a description of the IVGDR in hot nuclei is
given by a study of the isovector response and the relevant energy weighted
sums within the quantum RPA-like approaches 
\cite{bert83}-\cite{ayik08} or the semiclassical kinetic theory 
\cite{kola97}-\cite{baco99}. In the present work, following the
ideology of the kinetic theory, we consider both temperature and relaxation
effects on the IVGDR characteristics. Our semiclassical kinetic approach
ignores both the shell and single particle spin effects. Nevertheless, it
seems to be quite instructive for an investigation of the averaged properties
of the many-body systems. In many cases, it allows us to obtain analytical
results and represent them in a transparent way. There are also some conceptual
advantages in the use of the kinetic theory. Kinetic approach involves the
temperature directly into the equations of motion for the distribution
function, e.g., the temperature is considered here as a dynamic variable. In
contrast, in quantum approaches the temperature appears after ensemble
smearing of the observable quantities and can not be attributed to the
equation of motion for the wave function. Moreover the kinetic approach can
be easily generalized to consider the relaxation (damping) processes by
introduce of the collision integral \cite{abkh59,bape91}. Note that a
similar extension of the RPA due to the involving of the coupling with
$2p-2h $ states provides only the fine structure of the giant multipole
resonance (GMR) \cite{wamb88,kasp97} and an additional smearing procedure
for the strength function has to be used to derive the corresponding
collisional width of the GMR. Note also that the non-collisional
fragmentation width of the GMR (spreading of the GMR over non-collective
$1p-1h$ excitations in the RPA) does not drive the system toward a thermal
equilibrium but rather indicates a redistribution of the particle-hole
excitations in the vicinity of the collective state \cite{brag95}. In the
kinetic approach, this mechanism is presented due to the Landau damping.

In what follows we use the kinetic approach to study both the
temperature and mass number dependencies of the averaged
characteristics of the IVGDR, such as centroid energies, width,
isospin symmetry energy, mass coefficient, energy weighted sum (EWS)
and EWS enhancement factor. The corresponding analysis within RPA
and beyond RPA requires a large amount of numerical calculations,
see e.g. \cite{behe03}, which not necessary provide a clear
understanding of above mentioned macroscopic features of the IVGDR.
Both the quantum RPA and the kinetic approaches use the effective
nucleon-nucleon interaction. We apply the effective Landau
interaction in nuclear interior and the macroscopic boundary
condition in surface region. The macroscopic boundary condition
includes the phenomenological surface tension coefficient and
thereby substitutes for an effective interaction within the surface
layer. Note that the effective interaction is usually not quite
well-defined near nuclear surface because of strong particle density
inhomogeneity in this case and the involving of the relevant
boundary conditions can be used to improve the description. Our goal
is also to study the conditions for zero- to first- sound transition
in presence of the velocity-dependent forces and the effect of the
thermal Landau damping \cite{kola97} on the low-energy tail of the
strength function under different temperature and relaxation
conditions. Similar problems were considered in Refs. 
\cite{diko99}-\cite{baco99} by neglecting the velocity-dependent
part of Landau forces and using the simplest boundary condition of
Steinwedel-Jensen (SJ) model for the wave number $k=\pi /2R_{0}$,
where $R_{0}$ is the nuclear radius. However, it is well known
\cite{stri83} that such kind of boundary condition do not allow the
correct description of the $A$-dependence of the IVGDR energy for
light nuclei. A prove of the modified boundary condition for the
isovector eigenmode for a finite Fermi-liquid drop plies an
important role in our consideration.

There are different theoretical approaches to describe the
temperature behavior of the IVGDR width. Below, we will restrict
ourselves by the collisional damping and the thermal Landau
spreading. The alternative approach is the thermal fluctuation model
(TFM) in the adiabatic coupling scheme \cite{orbo96} which explains
the temperature increase of the IVGDR width as an effect of the
adiabatic coupling of the IVGDR to thermal shape fluctuations. One
can expect that the combine of both approaches will be able to
provide a satisfactory description of both the temperature and the
mass number dependencies of the IVGDR width \cite{gethorm98}. This
is an object of our forthcoming investigation.

Finally, we would like to note that we use the IVGDR as an
instrument to study the isovector motion in a spherical nuclear
Fermi-liquid drop. Our
claim is to describe the general features of collective excitations, such as
the $A$-dependence of the IVGDR energy and the isovector sound mode in nuclear
Fermi liquid ignoring many quantum effects. In particular, our approach
provides the possibility to compare the predictions of the standard liquid
drop model with the Fermi liquid drop one where the dynamic distortions of
Fermi surface is taking into account.

In Section 2 we apply the Landau's kinetic theory to the evaluation of the
response function in a two-component nuclear Fermi liquid. Both the
temperature and the relaxation phenomena are taken into account. In Section
3 we derive the energy weighted sums as the moments of the strength
function. Numerical illustrations, summary and conclusions are given in
Sections 4 and 5.

\section{Response function within the kinetic theory}

To derive the energy-weighted sums for the isovector excitations, we will
consider the density-density response of two-component nuclear matter to the
following external field

\begin{equation}
U_{\mathrm{ext}}(t)=\lambda _{0}e^{-i\omega t}\widehat{q}+\lambda _{0}^{\ast}
e^{i\omega t}\widehat{q}^{\ast },
\label{uext1}
\end{equation}
where $\lambda_{0}$ is the small amplitude, $\widehat{q}$ is the one-body
operator
\[
\widehat{q}=\sum\limits_{j=1}^{A}\widehat{q}(\vec{r}_{j},\xi
_{j})=\sum\limits_{j=1}^{A}\xi _{j}e^{-i\vec{q}\cdot \vec{r}_{j}}
\]%
and $\xi_{j}$ is the isotopic index ($\xi =p$ (or $+1$) for proton and
$\xi=n$ (or $-1$) for neutron). The response density-density function
$\chi(\omega )$ is given by \cite{ll1}
\begin{equation}
\chi (\omega )=\frac{\left\langle e^{-i\vec{q}\cdot \vec{r}}\right\rangle }{%
\lambda _{0}e^{-i\omega t}}=\frac{1}{\lambda _{0}e^{-i\omega t}}\int d\vec{r}%
\ e^{-i\vec{q}\cdot \vec{r}}\ \delta \rho _{\_}(\vec{r},t),  \label{chi1}
\end{equation}
where the isovector particle density variation $\delta \rho _{\_}(\vec{r}%
,t)\equiv \delta \rho _{\_}=\delta \rho _{n}-\delta \rho _{p}$ is due to the
external field $U_{\mathrm{ext}}(t)$ of Eq. (\ref{uext1}).

Below we will apply the kinetic theory in $(\vec{r},\vec{p})$ phase space to
the evaluation of the response function of Eq. (\ref{chi1}). The particle
density variation $\delta \rho _{\_}$ is then given by
\begin{equation}
\delta \rho _{\_}(\vec{r},t)=\int \frac{gd\vec{p}}{(2\pi \hbar )^{3}}\
\delta f_{\_}(\vec{r},\vec{p};t).  \label{rho1}
\end{equation}%
Here, $g=2$ is the spin degeneracy factor and $\delta f_{\_}(\vec{r},\vec{p}%
;t)\equiv \delta f_{\_}=\delta f_{n}-\delta f_{p}$ is the isovector
deviation of the distribution function $f(\vec{r},\vec{p};t)$ from the
equilibrium one at certain temperature $T$
\begin{equation}
f_{\mathrm{eq}}(\epsilon _{p})=\left[ 1+\exp {\frac{\epsilon _{p}-\mu }{T}}%
\right] ^{-1},  \label{feq}
\end{equation}%
where $\mu $ and $\epsilon _{p}=p^{2}/2m^{\ast }$ are respectively the
quasiparticle chemical potential and energy, $m^{\ast }$ is the effective
nucleonic mass.

A small isovector variation of the distribution function $\delta f_{\_}$ in
Eq. (\ref{chi1}) can be evaluated using the linearized kinetic equation. In
the nuclear volume, where inhomogeneity of the particle density is small,
the quasiparticle concept of the Landau-Fermi-liquid theory \cite{lipi} can
be justified. To evaluate $\delta f_{\_}$\ \ we will \ apply the linearized
Landau-Vlasov equation, completed by a source term $\delta St(f)$ for
relaxation processes in the following form \cite{abkh59,bape91}
\begin{equation}
\frac{\partial }{\partial t}\delta f_{\_}+\vec{v}\cdot \nabla _{\vec{r}%
}\delta f_{\_}-\nabla _{\vec{r}}(\delta U_{\mathrm{self}}+U_{\mathrm{ext}%
})\cdot \nabla _{\vec{p}}f_{\mathrm{eq}}=\mathrm{\delta St}[f],  \label{eq1}
\end{equation}%
where $\vec{v}=\vec{p}/m^{\ast }$ is the quasiparticle velocity. We point out that the
left hand side of kinetic equation (\ref{eq1}) can be derived \ by the
Wigner transformation from the linearized time dependent Hartree-Fock
equation\ in presence of the external field $U_{\mathrm{ext}}$ \cite{kota81}%
. The variation of the selfconsistent mean field $\delta U_{\mathrm{self}}$
in Eq. (\ref{eq1}) is then given by the Wigner transformation to the
corresponding mean field of the RPA. The selfconsistent mean field $\delta
U_{\mathrm{self}}$ is related to the Skyrme or Landau effective interaction
\cite{lilu91}. Within the Landau-Fermi-liquid theory the quantity $\delta U_{%
\mathrm{self}}$\ can be derived in terms of the Landau's interaction
amplitude $v_{\mathrm{int}}(\vec{p},\vec{p}^{\prime })$ as \cite%
{abkh59,bape91,brsy70}
\begin{equation}
\delta U_{\mathrm{self}}=\int \,{\frac{g\,d\vec{p}^{\prime }}{N_{T}\ (2\pi
\hbar )^{3}}}\,v_{\mathrm{int}}(\vec{p},\vec{p}^{\prime })\,\delta f_{\_}(%
\vec{r},\vec{p}^{\prime };t).  \label{self1}
\end{equation}%
The interaction amplitude $v_{\mathrm{int}}(\vec{p},\vec{p}^{\prime })$ is
parameterized in terms of the Landau constants $F_{l}^{\prime }$ as
\begin{equation}
v_{\mathrm{int}}(\vec{p},\vec{p}^{\prime })=\sum_{l=0}^{\infty
}\,F_{l}^{\prime }\,P_{l}(\hat{p}\cdot \hat{p}^{\prime }),\qquad \hat{p}=%
\vec{p}/p  \label{int1}
\end{equation}%
and the thermally averaged density of states $N_{T}$ in Eq. (\ref{self1}) is
introduced to provide the dimensionless constants $F_{l}^{\prime }$ in Eq. (%
\ref{int1}). Namely,
\begin{equation}
N_{T}=-\int \,{\frac{gd\vec{p}}{(2\pi \hbar )^{3}}}\,{\frac{\partial f_{%
\mathrm{eq}}(\epsilon _{p})}{\partial {\epsilon _{p}}}},  \label{nT}
\end{equation}%
with $N_{0}=g\,p_{F}\,m^{\ast }/2\,\pi ^{2}\,\hbar ^{3}$, where $p_{F}$ is
the Fermi momentum.

The right-hand side $\mathrm{\delta St}[f]$ of Eq. (\ref{eq1}) is the
Uehling-Uhlenbeck type collision integral linearized in $\delta f_{\_}$. The
collision integral $\mathrm{\delta St}[f]$ depends on the transition
probability of the two-nucleon scattering with initial momenta ($\vec{p}_{1},%
\vec{p}_{2}$) \ and final momenta ($\vec{p}_{1}^{\prime },\vec{p}%
_{2}^{\prime }$). At low temperatures $T\ll \epsilon _{F}$, where $\epsilon
_{F}$ is the Fermi energy, the momenta ($\vec{p}_{1},\vec{p}_{2}$)\ and ($%
\vec{p}_{1}^{\prime },\vec{p}_{2}^{\prime }$) are localized near the Fermi
surface and the relaxation time approximation can be used, see Refs. \cite%
{abkh59,bape91,lipi},
\begin{equation}
\mathrm{\delta St}[f]=-\sum_{lm,l\geq 1}^{\infty }\frac{1}{\tau _{l}}\delta
f_{\_,lm},  \label{st1}
\end{equation}%
where $\tau _{l}$ is the collisional relaxation time and $\delta f_{\_,lm}$
is a component of the $l,m$ multipolarity in $\vec{p}$-space of the
isovector variation $\delta f_{\_}$. Below we will restrict ourselves to the
azimuthally symmetric case (longitudinal perturbation) where $\delta f_{\_}$
depends only on the angle $\theta _{pq}$ between $\vec{p}$ and $\vec{q}$,
i.e. $\delta f_{\_,lm}$ is $m$-independent, see comment after Eq. (\ref{deltaf1}).
The partial relaxation time $%
\tau _{l}$ in Eq. (\ref{st1}), which corresponds to the Fermi-surface
distortion of multipolarity $l$, is derived in this case as \cite{brsy70}%
\begin{equation}
\frac{1}{\tau_{l}}=-\frac{\displaystyle\int_{0}^{\infty }d\epsilon _{p}\int d\Omega
_{p}Y_{l0}^{\ast }(\hat{p})\ \mathrm{\delta St}[f]}{\displaystyle\int_{0}^{\infty
}d\epsilon _{p}\int d\Omega _{p}Y_{l0}^{\ast }(\hat{p})\ \delta f_{\_}}.
\label{st2}
\end{equation}

Note that there is no term with $l=0$ in the sum (\ref{st1}) because of the
conservation relation for the particle number in collision processes.\ In
contrast to the case of isoscalar mode, the inclusion of the $l=1$ term in
the collision integral of Eq. (\ref{st1}) is due to the nonconservation of
the isovector current, i.e. due to the $pn$ collisions for the
counterstreaming neutron and proton flows. The numerical analysis shows \cite%
{diko99,laca99} that the\ isovector relaxation time $\tau _{l}$ depends only
slightly on the multipolarity $l\geq 2$ and we will use below the following
form for the collision integral
\begin{equation}
\mathrm{\delta St}[f]=-{\frac{1}{\tau _{1}}}\delta f_{\_}|_{l=1}-{\frac{1}{%
\tau _{2}}}\delta f_{\_}|_{l\geq 2}~.  \label{st3}
\end{equation}%
Here the notations $l=1$ and \ $l\geq 2$ mean that the perturbation of $%
\delta f_{\_}|_{l=1}$ and $\delta f_{\_}|_{l\geq 2}$ in the collision
integral includes only Fermi surface distortions with a multipolarity $l=1$
and $l\geq 2$, respectively.

The collisional relaxation time $\tau _{l}$ in Eq. (\ref{st2}) is
temperature and frequency dependent. The temperature dependence of $\tau _{l}
$ arises from the smeared out behavior of the equilibrium distribution
function $f_{\mathrm{eq}}$, see Eq. (\ref{feq}), near the Fermi momentum
\cite{abkh59,lipi}. The frequency dependence of $\tau _{l}$ is caused by the
memory (non-Markovian) effect in the collision integral. It can be shown,
see Ch. 8 of Ref. \cite{lipi}, that the presence of fast collective mode
changes the energy conservation factor in the collision integral $\mathrm{%
\delta St}[f]$ and provides the frequency dependence of the collisional
relaxation time $\tau _{l}$. Following Landau's prescription \cite{lipi}, we
will assume, see also Refs. \cite{diko99,laca99,ayyi98,kopl96},
\begin{equation}
\tau _{l}=\frac{\hbar \alpha _{l}}{T^{2}+(\hbar \omega /2\pi )^{2}}.
\label{tau1}
\end{equation}%
The parameter $\alpha _{l}$ in Eq. (\ref{tau1}) depends on the $NN$%
-scattering cross sections. In the case of isotropic energy independent
cross sections the result for $\alpha _{1}$ and $\alpha _{2}$ reads \cite%
{kopl96,aybo92}%
\begin{equation}
\alpha _{1}=3\,\epsilon _{F}^{2}/4\,\pi ^{2}\,\hbar \,\rho _{\mathrm{eq}%
}\,v_{F}\,\,\sigma _{-},\ \quad \alpha _{2}=5\,\epsilon _{F}^{2}/4\,\pi
^{2}\,\hbar \,\rho _{\mathrm{eq}}\,v_{F}\,\sigma _{\mathrm{av}},
\label{alpha1}
\end{equation}%
where $\epsilon _{F}$ is the Fermi energy, $v_{F}=p_{F}\,/m^{\ast }$ and $%
\rho _{\mathrm{eq}}$ is the bulk density in the nuclear interior. The $NN$%
-scattering cross sections $\sigma _{\mathrm{av}}$ and $\sigma _{-}$ in Eq. (%
\ref{alpha1}) are given by $\sigma _{\mathrm{av}}=(\sigma _{\mathrm{pp}%
}+\sigma _{\mathrm{nn}}+2\,\sigma _{\mathrm{pn}})/4$ and $\sigma _{-}=\sigma
_{\mathrm{pn}}/2$, where $\sigma _{\mathrm{pp}},\,\sigma _{\mathrm{nn}}$ and
$\sigma _{\mathrm{pn}}$ are the cross sections for nucleon pairs with
relative kinetic energy close to the Fermi energy. The value of $\alpha _{l}$
is significantly different for both vacuum and in-medium reduced cross
sections. Using the vacuum $NN$ cross sections \cite{lima93} $\sigma _{%
\mathrm{pp}}=\sigma _{\mathrm{nn}}=2.5\div 2.7\,\mathrm{fm}^{2}\,\,$and$%
\,\,\sigma _{\mathrm{pn}}=\sigma _{\mathrm{np}}=4.8\div 5.0\,\mathrm{fm}^{2}$%
, one obtains the following estimate of$\,\,\alpha _{2,\mathrm{vac}}=2.2\div
2.3\,\ \mathrm{MeV}$. The vacuum cross section is more appropriate in the
surface layer of the nucleus. Due to the Pauli blocking effect one can
expect that the $NN$ cross sections in nuclear matter should be lower than
the one in free space. Unfortunately, there is a strong uncertainty in the
derivation of the in-medium reduced $NN$ cross sections \cite{lima93}. We
will use the the following in-medium estimate of $\alpha _{2,\mathrm{bulk}}=$
$5.4\,\ \mathrm{MeV}$, see Refs. \cite{diko99,kopl96}.

In general the partial relaxation time $\tau _{2}$ in Eq. (\ref{st3}) is
larger than $\tau _{1}$. It is convenient to introduce the relation $\alpha
_{1}=\alpha _{2}/(1-\eta )$, where $\eta $ is the dimensionless parameter.
In the case of $\eta \rightarrow 1$ and $\alpha _{1}\rightarrow \infty $,
the relative motion of the proton-neutron fluids is not damped. The
character of damping of the isovector mode depends on the sign of parameter $%
\eta $. The zero-to-first sound transition is only possible for $\eta >0$
\cite{laca99}. For $\eta <0$, the relaxation due to $\tau _{1}$ leads to the
faster equilibration of the out phase proton-neutron motion than the
transition to the first sound.

At low temperatures only $\vec{p}$\ near the Fermi surface enter $\delta
f_{\_}$ and the solution of Eq. (\ref{eq1}) can be found in the form
\begin{equation}
\delta f_{\_}(\vec{r},\vec{p};t)=-{\frac{\partial f_{\mathrm{eq}}}{\partial
\epsilon _{p}}}\nu _{\omega ,\vec{q}}(\vec{p})\,e^{i(\vec{q}\cdot \vec{r}%
-\omega t)},  \label{deltaf1}
\end{equation}%
where $\partial f_{\mathrm{eq}}/\partial \epsilon _{p}$ is a sharply peaked
function at $p=p_{F}$ and $\nu _{\omega ,\vec{q}}(\vec{p})$ depends only on
the direction of $\vec{p}$. Moreover we consider the azimuthally symmetric
case where $\nu _{\omega ,\vec{q}}(\vec{p})$ depends only on the angle $%
\theta _{pq}$ between $\vec{p}$ and $\vec{q}$, and expand $\nu _{\omega ,%
\vec{q}}(\vec{p})$ in Legendre polynomials as
\begin{equation}
\nu _{\omega ,\vec{q}}(\vec{p})=\sum_{l=0}^{\infty }P_{l}(\cos \theta
_{pq})\,\nu _{l}(p).  \label{deltaf2}
\end{equation}

Using Eqs. (\ref{eq1}), (\ref{self1}), (\ref{deltaf1}) and (\ref{deltaf2}),
we obtain
\[
\left[ \left( \omega +\frac{i}{\tau _{2}}\right) -\vec{q}\cdot \vec{v}\right]
\nu _{\omega ,\vec{q}}(\vec{p})+\vec{q}\cdot \vec{v}\ \frac{1}{N_{T}}\int
\frac{gd\vec{p}^{\prime }}{(2\pi \hbar )^{3}}\ v_{\mathrm{int}}(\vec{p},\vec{%
p}^{\prime })\ \frac{\partial f_{\mathrm{eq}}}{\partial \epsilon _{p}}\ \nu
_{\omega ,\vec{q}}(\vec{p}^{\prime })
\]%
\begin{equation}
+\lambda _{0}\vec{q}\cdot \vec{v}=\frac{i}{\tau _{2}}\nu _{0}(p)P_{0}(\cos
\theta _{pq})+\eta \frac{i}{\tau _{2}}\nu _{1}(p)P_{1}(\cos \theta _{pq}).
\label{eq2}
\end{equation}%
Substituting expressions (\ref{int1}) and (\ref{deltaf2}) into Eq. (\ref{eq2}%
) and performing integration in Eq. (\ref{eq2}) over $\vec{p}$, one can come
to the following set of equations for the amplitudes $\nu _{l}(p)$:
\begin{eqnarray}
&&\nu _{l}(p)+(2l+1)\sum\limits_{l^{\prime }=0}^{\infty }\frac{F_{l^{\prime
}}^{\prime }}{2l^{\prime }+1}\widetilde{\nu }_{l^{\prime }}Q_{ll^{\prime
}}(z)-\lambda _{0}(2l+1)Q_{l0}(z)  \nonumber \\
&=&i(2l+1)\gamma \nu _{0}(p)\frac{1}{z}\left[ \delta _{l0}-Q_{l0}(z)\right]
-i(2l+1)\eta \gamma \nu _{1}(p)Q_{l0}(z).  \label{eq3}
\end{eqnarray}%
Here, $\widetilde{\nu }_{l}$ is the averaged amplitude

\begin{equation}
\widetilde{\nu }_{l}=-\frac{1}{N_{T}}\int \frac{gd\vec{p}}{(2\pi \hbar )^{3}}%
\frac{\partial f_{\mathrm{eq}}(\epsilon _{p})}{\partial \epsilon _{p}}\ \nu
_{l}(p),  \label{nul1}
\end{equation}%
and
\[
Q_{ll^{\prime }}(z)=-\frac{1}{2}\int\limits_{-1}^{1}dx\frac{P_{l}(x)\ x\
P_{l^{\prime }}(x)}{z-x}, \quad
x=\cos \theta _{pq},\quad \gamma =\frac{1}{\tau _{2}qv},\quad z=s+i\gamma
,\quad s=\frac{\omega }{qv}.
\]%
For simplicity we will assume
\begin{equation}
F_{l=0}^{\prime }\neq 0,\,\,\,\,\,\,\,F_{l=1}^{\prime }\neq 0\,,\quad
\,F_{\ell \geq 2}^{\prime }=0~.  \label{int2}
\end{equation}%
Under the condition (\ref{int2}), the basic equations (\ref{eq3}) can be
solved with respect to the amplitude $\widetilde{\nu }_{0}$. After rather
simple calculation one can obtain (see \textrm{Appendix A})%
\begin{equation}
\widetilde{\nu }_{0}=\frac{\widetilde{\chi }_{\mathrm{in}}(\omega ,q)}{%
1+F_{0}^{\prime }\ \widetilde{\chi }_{\mathrm{in}}(\omega ,q)}\lambda _{0},
\label{nu0}
\end{equation}%
where the internal response function $\widetilde{\chi }_{\mathrm{in}}(\omega
,q)$ is given by Eq. (\ref{A6}).

The amplitude $\widetilde{\nu }_{0}$ is related to the density-density
response function of Eq. (\ref{chi1}). Substituting Eq. (\ref{deltaf1}) into
Eq. (\ref{rho1}) and using Eq. (\ref{deltaf2}), one obtains%
\begin{equation}
\delta \rho _{-}(\vec{r},t)=-\int \frac{gd\vec{p}}{(2\pi \hbar )^{3}}{\frac{%
\partial f_{\mathrm{eq}}(\epsilon _{p})}{\partial \epsilon _{p}}}\ \nu
_{0}(p)\ e^{i(\vec{q}\cdot \vec{r}-\omega t)}.  \label{rho2}
\end{equation}%
Using definition (\ref{chi1}) and Eqs. (\ref{nul1}) and (\ref{rho2}), we
obtain the density-density response function $\chi (\omega ,q)$ for a given
momentum transfer $q$ in the following form
\begin{equation}
\chi (\omega ,q)=\frac{2\ N_{T}\ \widetilde{\chi }_{\mathrm{in}}(\omega ,q)}{%
1+F_{0}^{\prime }\ \widetilde{\chi }_{\mathrm{in}}(\omega ,q)}.  \label{chi2}
\end{equation}%
Equation (\ref{chi2}) (together with (\ref{A6})) gives a generalization of
analogous result of Refs. \cite{diko99,laca99} to the case of the velocity
dependent (nonlocal) interaction $F_{1}^{\prime }\neq 0$. The poles of the
response function $\chi (\omega ,q)$ give the eigenfrequencies of collective
excitations $\omega =\omega _{R}+i\omega _{I}$ and satisfy the following
dispersion relation
\begin{equation}
1+F_{0}^{\prime }\ \widetilde{\chi }_{\mathrm{in}}(\omega ,q)=0.
\label{disp1}
\end{equation}

\subsection{Boundary condition}

For finite nuclei, the dispersion relation (\ref{disp1}) has to be augmented
by the boundary condition. The boundary condition can be taken as a
condition for the balance of the forces on the free nuclear surface
\begin{equation}
\vec{n}\cdot \vec{F|}_{S}+\vec{n}\cdot \vec{F}_{S}=0,  \label{B1}
\end{equation}%
where $\vec{n}$ is the unit vector in the normal direction to the nuclear
surface $S$, the internal force $\vec{F}$ is associated with the isovector
sound wave and $\vec{F}_{S}$ is the isovector surface tension force. The
internal force $\vec{F}$ is derived by the momentum flux tensor\ in the
nuclear interior and can be evaluating directly using the basic kinetic
equation (\ref{eq1}), see Appendix B. The isovector surface force $\vec{F}%
_{S}$ is due to the isovector polarization at the nuclear surface \cite%
{mysw74}. Both forces $\vec{n}\cdot \vec{F|}_{S}$ and $\vec{n}\cdot \vec{F}%
_{S}$ in Eq. (\ref{B1}) can be represented in terms of isovector shift of
the nuclear surface and the boundary condition (\ref{B1}) takes the final
form of the following secular equation the wave number $q$, see Appendix B,
\begin{equation}
\left[ \frac{\bar{\rho}_{\mathrm{eq}}}{4}C_{\mathrm{sym}}+\frac{\mu _{F}}{3}-%
\frac{\mu _{F}}{x^{2}}\right] j_{1}(x)+\left[ \frac{\mu _{F}}{x}-\frac{2\rho
_{\mathrm{eq}}Q_{\mathrm{sym}}}{3qr_{0}(1+\kappa _{NM})}\right]
j_{1}^{\prime }(x)=0,  \label{seceq}
\end{equation}%
where $x=qR_{0}$, $\ R_{0}=r_{0}A^{1/3}$, the $\ $parameter $\mu _{F}$ \
derives the Fermi-surface distortion effect, see Eq. (\ref{B6}) in Appendix
B, and $Q_{\mathrm{sym}}$ is the isovector surface tension coefficient, see
Eq. (\ref{B10}). In the limit $Q_{\mathrm{sym}}\rightarrow \infty $, the
boundary condition (\ref{seceq}) leads to the boundary condition $%
j_{1}^{\prime }(x)=0$ of the Steinwedel-Jensen model \cite{bomo75}. We point
out that the secular equation (\ref{seceq}) for $q$ has to be solved
consistently with the dispersion relation (\ref{disp1}) for $s$.

\section{Energy-weighted sums and transport coefficients}

The presence of the nonlocal interaction in Eq. (\ref{chi2}) leads to some
important consequences for the properties of the EWS $m_{k}(q)$ for
isovector mode. Let us introduce the strength function per unit volume
\begin{equation}
S(\omega ,q)=\mathrm{Im}\chi (\omega ,q)/\pi .  \label{str1}
\end{equation}%
The EWS are defined by
\begin{equation}
m_{k}(q)=\int\limits_{0}^{\infty }d(\hbar \omega )\ (\hbar \omega )^{k}\
S(\omega ,q).  \label{mk1}
\end{equation}%
Note that the EWS $m_{k}(q)$\ for odd $k$ can also be evaluated by use the
dynamic polarizability $\mathrm{Re}\chi (\omega ,q)$. Considering the two
limits $\omega \rightarrow 0$ and $\omega \rightarrow \infty $ one can
obtain \cite{list89}
\begin{equation}
\left. \mathrm{Re}\chi (\omega ,q)\right\vert _{\omega \rightarrow 0}=2\left[
m_{-1}(q)+(\hbar \omega )^{2}m_{-3}(q)+...\right]  \label{mk2}
\end{equation}%
and
\begin{equation}
\left. \mathrm{Re}\chi (\omega ,q)\right\vert _{\omega \rightarrow \infty }=-%
\frac{2}{(\hbar \omega )^{2}}\left[ m_{1}(q)+(\hbar \omega )^{-2}m_{3}(q)+...%
\right] .  \label{mk3}
\end{equation}%
In the case of cold nucleus $T=0$ and no relaxation $\tau _{1},\tau
_{2}\rightarrow \infty $, applying Eqs. (\ref{mk2}) and (\ref{mk3}) to the
response function (\ref{chi2}) and using Eqs. (\ref{A6})-(\ref{A10}), we
recover well-known results \cite{list89}%
\begin{equation}
m_{-1}^{(0)}(q)=\frac{\rho _{\mathrm{eq}}}{2\ C_{\mathrm{sym}}},\quad
m_{1}^{(0)}(q)=\hbar ^{2}\frac{\rho _{\mathrm{eq}}}{2\ m^{\prime }}%
q^{2},\quad m_{3}^{(0)}(q)=\hbar ^{4}\frac{C_{\mathrm{sym}}^{\prime }\ \rho
_{\mathrm{eq}}}{2\ m^{\prime 2}}q^{4}.  \label{mk4}
\end{equation}%
Here, $\rho _{\mathrm{eq}}$ is the equilibrium particle density, $C_{\mathrm{%
sym}}$ is the isospin symmetry energy

\begin{equation}
C_{\mathrm{sym}}=b_{\mathrm{sym,vol}}=\frac{2}{3}\epsilon
_{F}(1+F_{0}^{\prime })\approx 60\ \mathrm{MeV,}  \label{c1}
\end{equation}%
where $b_{\mathrm{sym,vol}}$ is the volume part of symmetry energy in the
nuclear mass formula \cite{bomo75}, $\epsilon _{F}$ is the Fermi energy and $%
m^{\prime }$ is the effective mass for isovector mode
\begin{equation}
m^{\prime }=\frac{m^{\ast }}{1+F_{1}^{\prime }/3},\qquad m^{\ast
}=m(1+F_{1}/3)  \label{mas1}
\end{equation}%
and the upper index \textquotedblright $(0)$\textquotedblright\ indicates
that the corresponding quantity is taken for $T=0$ with $\tau _{1},\tau
_{2}\rightarrow \infty $. The renormalized symmetry energy $C_{\mathrm{sym}%
}^{\prime }$ in Eq. (\ref{mk4}) is given by
\begin{equation}
C_{\mathrm{sym}}^{\prime }=C_{\mathrm{sym}}+8\ \epsilon _{F}/15,  \label{c2}
\end{equation}%
where the last term on the r.h.s. is due to the Fermi surface distortion
effect \cite{kosh01}.

In contrast to the isoscalar mode, the isovector EWS $m_{1}(q)$ of Eq. (\ref%
{mk4}) is model dependent. As can be seen from Eq. (\ref{mk4}), the sum $%
m_{1}(q)$ includes the enhancement factor (for nuclear matter) \cite{krtr80}%
\begin{equation}
1+\kappa _{NM}=(m/m^{\ast })(1+F_{1}^{\prime }/3),  \label{kappa1}
\end{equation}%
which depends on the nonlocal interaction constant $F_{1}^{\prime }\neq 0$.

In general the inverse EWS $m_{-1}$ derives the static stiffness
coefficient. Evaluating the distorted wave function $\left\vert \Psi _{%
\mathrm{ad}}\right\rangle $ for a static (adiabatic) constrained field $U_{%
\mathrm{ext}}=\lambda _{0}\widehat{q}+\lambda _{0}^{\ast }\widehat{q}^{\ast
} $\ (see Eq. (\ref{uext1}) at $\omega \longrightarrow 0$), one can evaluate
the corresponding variation of the energy
\begin{equation}
\delta E=\left\langle \Psi _{\mathrm{ad}}\right\vert \widehat{H}\left\vert
\Psi _{\mathrm{ad}}\right\rangle -\left\langle \Psi _{\mathrm{eq}%
}\right\vert \widehat{H}\left\vert \Psi _{\mathrm{eq}}\right\rangle =\frac{1%
}{4\ m_{-1}^{(0)}}\delta Q^{2},  \label{de1}
\end{equation}%
where $\widehat{H}$ is the non-perturbed Hamiltonian of the nucleus and $%
\delta Q$ is the change of the mean value $\left\langle \widehat{q}%
\right\rangle $ induced by the distorted wave function. Using Eqs. (\ref{mk4}%
) and (\ref{de1}), we obtain the adiabatic stiffness coefficient (per unit
volume) $C_{Q,\mathrm{ad}}$ in the following form%
\begin{equation}
C_{Q,\mathrm{ad}}=\frac{\partial ^{2}\delta E}{\partial \ \delta Q^{2}}=%
\frac{1}{2m_{-1}^{(0)}}=C_{\mathrm{sym}}/\rho _{\mathrm{eq}}.  \label{cq1}
\end{equation}

The cubic sum $m_{3}$ is related to the stiffness coefficient $C_{Q,\mathrm{%
sc}}$ of the scaling approximation \cite{bola79}. Assuming a scaled form of
the perturbed ground state wave function $\left\vert \Psi _{\mathrm{sc}%
}\right\rangle =e^{-i\lambda _{0}\widehat{q}}\left\vert \Psi _{\mathrm{eq}%
}\right\rangle $, one can evaluate the corresponding change of the energy
\begin{equation}
\delta E=\left\langle \Psi _{\mathrm{sc}}\right\vert \widehat{H}\left\vert
\Psi _{\mathrm{sc}}\right\rangle -\left\langle \Psi _{\mathrm{eq}%
}\right\vert \widehat{H}\left\vert \Psi _{\mathrm{eq}}\right\rangle =\frac{%
m_{3}^{(0)}}{4m_{1}^{(0)2}}\delta Q^{2}.  \label{de2}
\end{equation}%
Using Eqs. (\ref{mk4}), the scaled stiffness coefficient $C_{Q,\mathrm{sc}}$
takes the following form
\begin{equation}
C_{Q,\mathrm{sc}}=\frac{m_{3}^{(0)}}{2m_{1}^{(0)2}}=C_{\mathrm{sym}}^{\prime
}/\rho _{\mathrm{eq}}.  \label{cq2}
\end{equation}%
Note that the constrained stiffness coefficient $C_{Q,\mathrm{ad}}$ of Eq. (%
\ref{cq1}) is not affected by the Fermi-surface distortion since the sum $%
m_{-1}$ contains the static symmetry energy $C_{\mathrm{sym}}$. This is not
case for the stiffness coefficient $C_{Q,\mathrm{sc}}$ because the
renormalized $C_{\mathrm{sym}}^{\prime }$ enters the sum $m_{3}$, see Eq. (%
\ref{mk4}). \ It can be shown \cite{kolu07}\ that both the sum $m_{3}$ and
the stiffness coefficient $C_{Q,\mathrm{sc}}$ depend on the Fermi-surface
distortions of multipolarity $l\leq 2$.

The EWS $m_{-1},\ m_{1}$ and $m_{3}$ allow obtaining the adiabatic, $%
\widetilde{E}_{\mathrm{ad}}$, and scaled, $\widetilde{E}_{\mathrm{sc}}$,
average energies (centroid energies) of IVGDR \cite{koko99}%
\begin{equation}
\widetilde{E}_{\mathrm{ad}}=\sqrt{\frac{m_{1}^{(0)}}{m_{-1}^{(0)}}}=\hbar
\sqrt{\frac{C_{\mathrm{sym}}}{m^{\prime }}}q,\qquad \widetilde{E}_{\mathrm{sc%
}}=\sqrt{\frac{m_{3}^{(0)}}{m_{1}^{(0)}}}=\hbar \sqrt{\frac{C_{\mathrm{sym}%
}^{\prime }}{m^{\prime }}}q.  \label{even1}
\end{equation}%
There is a significant difference between the adiabatic energy, $\widetilde{E%
}_{\mathrm{ad}}$, i.e., derived by\ a static stiffness coefficient $C_{%
\mathrm{sym}}$, and the scaled one, $\widetilde{E}_{\mathrm{sc}}$,
associated with the isovector sound in nuclear Fermi liquid. The
Fermi-surface distortion effects increase the stiffness coefficient $C_{%
\mathrm{sym}}^{\prime }$ Eq. (\ref{c2}) and leads to an increase of the
centroid energy $\widetilde{E}_{\mathrm{sc}}$ of the isovector mode.

The low frequency (see Eq. (\ref{mk2})) sum $m_{-3}^{(0)}$ is related to the
transport coefficient. Assuming a slow time dependence of the external field
$U_{\mathrm{ext}}(t)$ and evaluating the corresponding time dependent wave
function $\left\vert \Psi (t)\right\rangle $, one can find the change of the
energy in the following form%
\begin{equation}
\delta E=\left\langle \Psi (t)\right\vert \widehat{H}\left\vert \Psi
(t)\right\rangle -\left\langle \Psi _{\mathrm{eq}}\right\vert \widehat{H}
\left\vert \Psi _{\mathrm{eq}}\right\rangle =\hbar ^{2}\frac{m_{-3}^{(0)}} {%
4m_{-1}^{(0)2}}\delta \dot{Q}^{2}.  \label{de3}
\end{equation}
Using Eq. (\ref{de3}) we obtain the transport (mass) coefficient $B_{Q}$ as
\begin{equation}
B_{Q}=\frac{\partial ^{2}\delta E}{\partial \ \delta \dot{Q}^{2}}= \hbar ^{2}%
\frac{m_{-3}^{(0)}}{2m_{-1}^{(0)2}}.  \label{b1}
\end{equation}
Using the mass coefficient $B_{Q}$ and the stiffness coefficient $C_{Q,%
\mathrm{ad}}$ (see Eq. (\ref{cq1})), the eigenfrequency $\omega_{\mathrm{macr%
}}$ and corresponding eigenenergy $E_{\mathrm{macr}}$ for the macroscopic
eigenvibrations can be derived as
\begin{equation}
\omega_{\mathrm{macr}}=\sqrt{\frac{C_{Q,\mathrm{ad}}}{B_{Q}}}, \qquad E_{%
\mathrm{macr}}=\hbar\omega_{\mathrm{macr}}=\sqrt{\frac{m_{-1}^{(0)}} {%
m_{-3}^{(0)}}}.  \label{eclas1}
\end{equation}
We point out that the sum $m_{-3}^{(0)}$ can not be evaluated within the
Landau-Vlasov kinetic approach by the use of the low frequency expansion of
Eq. (\ref{mk2}). That is because the Fermi liquid gets into Landau-damping
regime at $\omega \sim 0$ and special attention must be paid to the low
energy region in Eq. (\ref{mk1}), see below.

\section{Results and Discussions}

\bigskip

We will present results of numerical calculations based on the consideration
of previous sections. In this work we adopt the value of $r_{0}=1.2\,\mathrm{%
fm}$ and the effective nucleon mass $m^{\ast }$ is taken as $m^{\ast }=0.9m$%
\textbf{\ }which corresponds to the Landau parameter $F_{1}=-\ 0.3$. For the
isovector interaction parameter $F_{0}^{\prime }$ we have used $%
F_{0}^{\prime }=1.41$ to keep a reasonable value of isospin symmetry energy $%
C_{\mathrm{sym}}$ of the order of $60\ \mathrm{MeV}$, see Eq. (\ref{c1}).
The interaction parameter $F_{1}^{\prime }$ will be derived and discussed
below. For the relaxation parameters in Eq. (\ref{tau1}) we use the values
of $\alpha _{2}=$ $5.4\,\ \mathrm{MeV}$ and $\eta =-\ 0.1$ which correspond
to the in-medium reduced $NN$\ cross sections \cite{diko99,laca99}. For more
clear interpretation of some numerical results we will also use the
relaxation parameters $\alpha _{2}$ and $\eta $\ beyond these
well-established values. \

\subsection{Eigenenergy and enhancement factor in cold nuclei}

The interaction parameter $F_{1}^{\prime }$ can be estimated by considering
the enhancement factor $1+\kappa _{NM}$ for the isovector EWS $m_{1}(q)$,
see Eq. (\ref{kappa1}). Following Ref. \cite{diko99}, we derive the
photoabsorption cross section $\sigma _{\mathrm{abs}}(\omega )$\ in terms of
the strength function $S(\omega ,q)$ Eq. (\ref{str1}) as follows%
\begin{equation}
\sigma _{\mathrm{abs}}(\omega )={\frac{4\pi ^{2}e^{2}}{cq^{2}\rho _{\mathrm{%
eq}}}}{\frac{NZ}{A}}\omega S(\omega ,q).  
\label{sigma1}
\end{equation}
In the case of the velocity independent $NN$-interaction, the cross section $%
\sigma _{\mathrm{abs}}(\omega )$ (\ref{sigma1}) is normalized by the
ordinary Thomas-Reiche-Kuhn sum rule \cite{risc80} (see $m_{1}(q)$ in Eq. (%
\ref{mk4}) for $\kappa _{NM}=0$)
\begin{equation}
\widetilde{m}_{1,\mathrm{TRK}}^{(0)}=\int\limits_{0}^{\infty }\ d(\hbar
\omega )\ \sigma _{\mathrm{abs}}(\omega )=\frac{2\pi ^{2}\hbar e^{2}}{mc}%
\frac{NZ}{A}\quad \textrm{for}\quad \kappa _{NM}=0  \label{mk5}
\end{equation}
at $T=0$ and $\tau _{1},\tau _{2}\rightarrow \infty $.

Taking into account the velocity dependence of the $NN$-interaction with $%
F_{1}\neq 0$ and $F_{1}^{\prime }\neq 0$, we note that both the enhancement
factor $\kappa _{NM}\neq 0$ in $m_{1}(q)$ of Eq. (\ref{mk4}) and the
corresponding correction at the last term of the boundary condition (\ref%
{B18}) affect the sum rule (\ref{mk5}). For $\kappa _{NM}\neq 0$, using Eqs.
(\ref{sigma1}), (\ref{mk4}) and (\ref{B18}), we obtain the following result
\cite{kolu07}
\begin{equation}
\widetilde{m}_{1}^{(0)}=\int\limits_{0}^{\infty }\ d(\hbar \omega )\
\sigma _{\mathrm{abs}}(\omega ) 
=\frac{2\pi ^{2}\hbar e^{2}}{mc}\left( \frac{q_{1}^{\prime }(A)}{q_{0}(A)}
\right) ^{2}\frac{NZ}{A}(1+\kappa_{NM})\quad \textrm{for}\quad \kappa_{NM}>0,
\label{mk6}
\end{equation}
at $T=0$ and $\tau _{1},\tau _{2}\rightarrow \infty $. Here $q_{0}(A)$ and $%
q_{1}^{\prime }(A)$ are the lowest roots of Eq. (\ref{B18}) for $\kappa
_{NM}=0$ and $\kappa _{NM}\neq 0$, respectively. The value of interaction
parameters $F_{1}^{\prime }$ can be now obtained from a fit of the evaluated
enhancement factor $\widetilde{m}_{1}^{(0)}/\widetilde{m}_{1,\mathrm{TRK}%
}^{(0)}$ to the experimental data. In this work we have adopted $%
F_{1}^{\prime }=1.1$. Our estimate of the enhancement factor is about 10\%
for light nuclei and increases to 20\% for heavy nuclei which is in a good
agreement with experimental data \cite{BeFu}.

In finite nuclei, both the IVGDR eigenenergy $E_{\mathrm{IVGDR}}$ and the
EWS $m_{k}$ are the complicated functions of the mass number $A$. In
contrast to the classical Steinwedel-Jensen model \cite{bomo75}, the value $%
E_{\mathrm{IVGDR}}\cdot A^{1/3}$ is not the constant but increases with $A$
\cite{BeFu}. Within our Fermi liquid approach, the $A$-dependence of the
IVGDR eigenenergy and EWS occurs due to the boundary condition of Eq. (\ref%
{seceq}) on the moving nuclear surface.

\begin{figure}
\begin{center}
\includegraphics*[width=9cm]{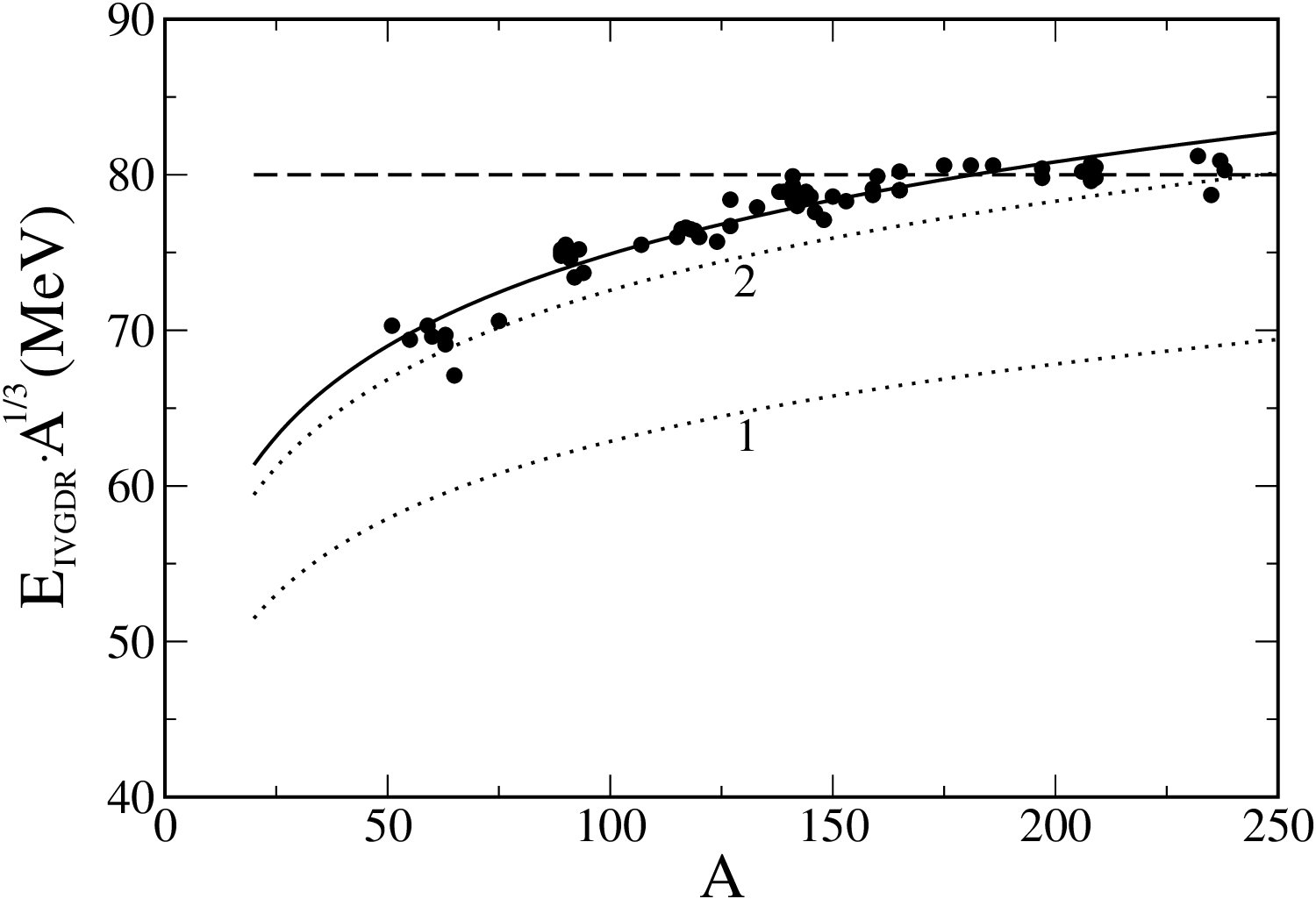}
\end{center}
\caption{Dependence of the IVGDR energy on the mass number $A$: the dotted line 1
is the first-sound regime without Fermi surface distortion (adiabatic approximation,
see $\widetilde{E}_{\mathrm{ad}}$ in the Eq. (\ref{even1})); the dotted line 2 shows
the calculation which takes into account the Fermi surface distortion up to the
multipolarity $l=2$ (scaling approximation, see $\widetilde{E}_{\mathrm{sc}}$ in the
Eq. (\ref{even1})); the solid line was obtained by use of the dispersion relation
(\ref{disp1}) and the secular equation (\ref{seceq}). The dashed line shows the result
within SJ model with commonly used \cite{ayik08,diko99,laca99} value of the wave number
$q=\pi /2R_{0}$. The solid dots are the experimental data from Ref. \cite{BeFu}.}
\label{fig1}
\end{figure}

\bigskip

In \figurename\ \ref{fig1} we show the dependence of the IVGDR energy (multiplied by $A^{1/3}$%
) on the mass number $A$. The calculations have been performed for $Q_{%
\mathrm{sym}}=10.5$ \textrm{MeV} and $F_{1}^{\prime }=1.1$. The solid line
is the eigenenergy obtained from the dispersion equation (\ref{disp1})
augmented by the boundary condition of Eq. (\ref{seceq}). Both dotted lines
in \figurename\ \ref{fig1} have been obtained from the EWS definitions of the centroid
energies Eq. (\ref{even1}) (curve 1 for $\widetilde{E}_{\mathrm{ad}}$ and
curve 2 for $\widetilde{E}_{\mathrm{sc}}$). The significant upward shift of
the scaled energy, $\widetilde{E}_{\mathrm{sc}}$, with respect to the
constrained one, $\widetilde{E}_{\mathrm{ad}}$, is due to the Fermi surface
distortions of multipolarity $l\leq 2$ presented in $\widetilde{E}_{\mathrm{%
sc}}$. An additional upward shift of the exact eigenenergy (solid line) is
due to the higher multipolarities $l>2$ of the Fermi surface distortions
presented in the dispersion equation (\ref{disp1}). As seen in \figurename\ \ref{fig1}, we
reproduce quite well the average behavior of the IVGDR energy $E_{\mathrm{%
IVGDR}}$. Note that the boundary condition $j_{1}^{\prime }(qR_{0})=0$ of
the Steinwedel-Jensen model \cite{bomo75} as well as the commonly used wave
number $q=\pi /2R_{0}$ \cite{ayik08,diko99,laca99} do not describe the $A$%
-dependence of \ the IVGDR energy correctly, see dashed line in \figurename\ \ref{fig1}.

\subsection{First sound regime}

We consider the first sound regime as the displacement of the
spherically-symmetric Fermi surface without its deformation in momentum
space. In the case of the velocity dependent effective $NN$-interaction with
$F_{1}^{\prime }\neq 0$, the first sound eigenvibrations differ from the
corresponding ones in the classical Steinwedel-Jensen model \cite{bomo75}.
For the sake of simplicity we will consider the longitudinal eigenvibrations
assuming $U_{\mathrm{ext}}(t)=0$ and no relaxation ($\tau _{k}=\infty $) in
Eq. (\ref{eq1}). Using Eq. (\ref{deltaf1}) and expansion (instead of Eq. (%
\ref{deltaf2}))
\begin{equation}
\nu _{\omega ,\vec{q}}(\vec{p})=\sum_{lm}\nu _{lm}(\vec{q},\omega )Y_{lm}(%
\hat{p}),  \label{nu1}
\end{equation}%
we will transform the kinetic equation (\ref{eq1}) to the following set of
equations%
\begin{equation}
\omega \ \nu _{lm}-v_{F}\ q\sum_{l^{\prime }m^{\prime }}G_{l^{\prime
}}^{\prime }\left\langle lm\right\vert \hat{q}\cdot \hat{p}\left\vert
l^{\prime }m^{\prime }\right\rangle \ \nu _{l^{\prime }m^{\prime }}=0,
\label{eq4}
\end{equation}%
where $G_{l}^{\prime }=1+F_{l}^{\prime }/(2l+1)$ and
\[
\left\langle lm\right\vert \hat{q}\cdot \hat{p}\left\vert l^{\prime
}m^{\prime }\right\rangle =\int d\Omega _{p}\ Y_{lm}^{\ast }(\hat{p})\cos
(\theta _{qp})Y_{l^{\prime }m^{\prime }}(\hat{p}).
\]%
Using Eq. (\ref{int2}) and assuming $\nu _{lm}|_{l\geq 2}=0$, we obtain from
Eq. (\ref{eq4}) the following\ closed equations for amplitudes $\nu _{00}$
and $\nu _{10}$:
\begin{equation}
s\ \nu _{00}-\frac{1}{\sqrt{3}}G_{1}^{\prime }\nu _{10}=0,\qquad s\ \nu
_{10}-\frac{1}{\sqrt{3}}G_{0}^{\prime }\nu _{00}=0  \label{eq5}
\end{equation}%
and the corresponding dispersion relation%
\begin{equation}
\omega =\frac{1}{\sqrt{3}}v_{F}\ q\sqrt{G_{0}^{\prime }\ G_{1}^{\prime }}.
\label{disp2}
\end{equation}%
Finally, using the definition (\ref{c1}) of the symmetry energy $C_{\mathrm{%
sym}}$, we obtain the eigenenergy energy $E_{\mathrm{first}}$ of the IVGDR
in the first sound limit in the following form%
\begin{equation}
E_{\mathrm{first}}=\hbar \sqrt{\frac{C_{\mathrm{sym}}}{m}}\frac{%
1+F_{1}^{\prime }/3}{1+F_{1}/3}q=\hbar \sqrt{\frac{C_{\mathrm{sym}}}{m}}%
(1+\kappa _{NM})\ q.  \label{eres1}
\end{equation}%
The energy $E_{\mathrm{first}}$ of Eq. (\ref{eres1}) differs from the one $%
E_{\mathrm{SJ}}$ of the Steinwedel-Jensen model
\begin{equation}
E_{\mathrm{SJ}}=\hbar \sqrt{\frac{C_{\mathrm{sym}}}{m}}q  \label{esj}
\end{equation}%
by the enhancement factor $1+\kappa _{NM}$. Similar result was reported
earlier in Ref. \cite{krtr80}.

\subsection{Strength function}

\begin{figure}
\begin{center}
\includegraphics*[width=9cm]{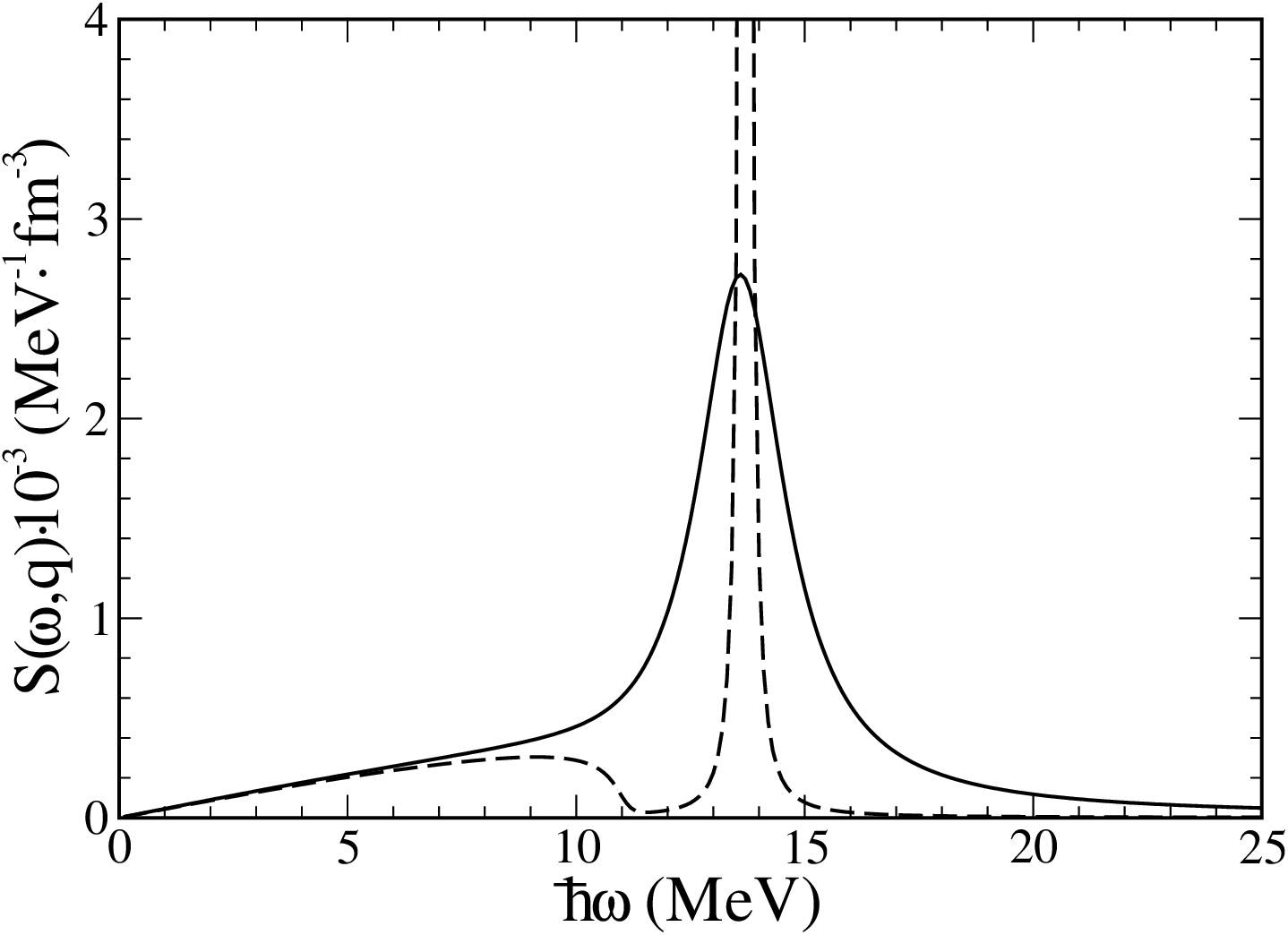}
\end{center}
\caption{Strength function $S(\omega,q)$ from Eqs. (\ref{chi2}) and (\ref{A10})
for $F_{1}=-\ 0.3$, $F_{1}^{\prime }=1.1$, $\eta =-\ 0.1$, $A=208$. Solid line
for $T=2$ MeV, $\alpha_{2}=5.4$ MeV and dashed line for $T=0.5$ MeV,
$\alpha_{2}=100$ MeV.}
\label{fig2}
\end{figure}

Performing the numerical calculations of the response function $\chi (\omega
,q)$ (\ref{chi2}), one can evaluate the strength function $S(\omega ,q)$ (%
\ref{str1}) and the EWS $m_{k}(q)$ (\ref{mk1}) for finite temperatures $%
T\neq 0$ and in presence of relaxation. The strength function $S(\omega ,q)$
is sensitive to the interaction parameters and to the relaxation properties.
Because of $F_{0}^{\prime }>0$, the IVGDR strength function contains both
the sound mode contribution at $s>1$ and the Landau damping region at $s<1$.
This is illustrated in \figurename\ \ref{fig2}.

\bigskip

To show the presence of the Landau damping in the IVGDR $S(\omega,q)$ in a
transparent way, we have plotted in \figurename\ \ref{fig2} the result for
the zero-sound regime $\omega _{R}\tau _{2}\gg 1$ at $T=0.5$ MeV, $\alpha_{2}=100$
MeV (dashed line). The Landau damping appears there as a wide bump on the left
side of the narrow sound peak. For high temperature (solid line in \figurename\
\ref{fig2}), the sound peak becomes wider due to the decrease of the relaxation
time (collisional relaxation), see Eq. (\ref{tau1}), and due to the collisionless
thermal Landau damping, which increases with $T$, see Ref. \cite{kola97}. As
can be seen from \figurename\ \ref{fig2}, overlapping of both the sound peak
and the Landau damping bump leads to the asymmetry of the IVGDR\ strength
function at high temperatures. This feature of the IVGDR\ strength function
is observed experimentally \cite{baum98}. Note that the IVGDR width, which can
be derived from \figurename\ \ref{fig2}, represents a collisional part of total
width only and this one is significantly smaller the experimental width of Ref.
\cite{baum98}. Additional part of the IVGDR width is caused by the fragmentation
mechanism and we will take into account this fact below in \figurename\ \ref{fig8}.

\begin{figure}
\begin{center}
\includegraphics*[width=9cm]{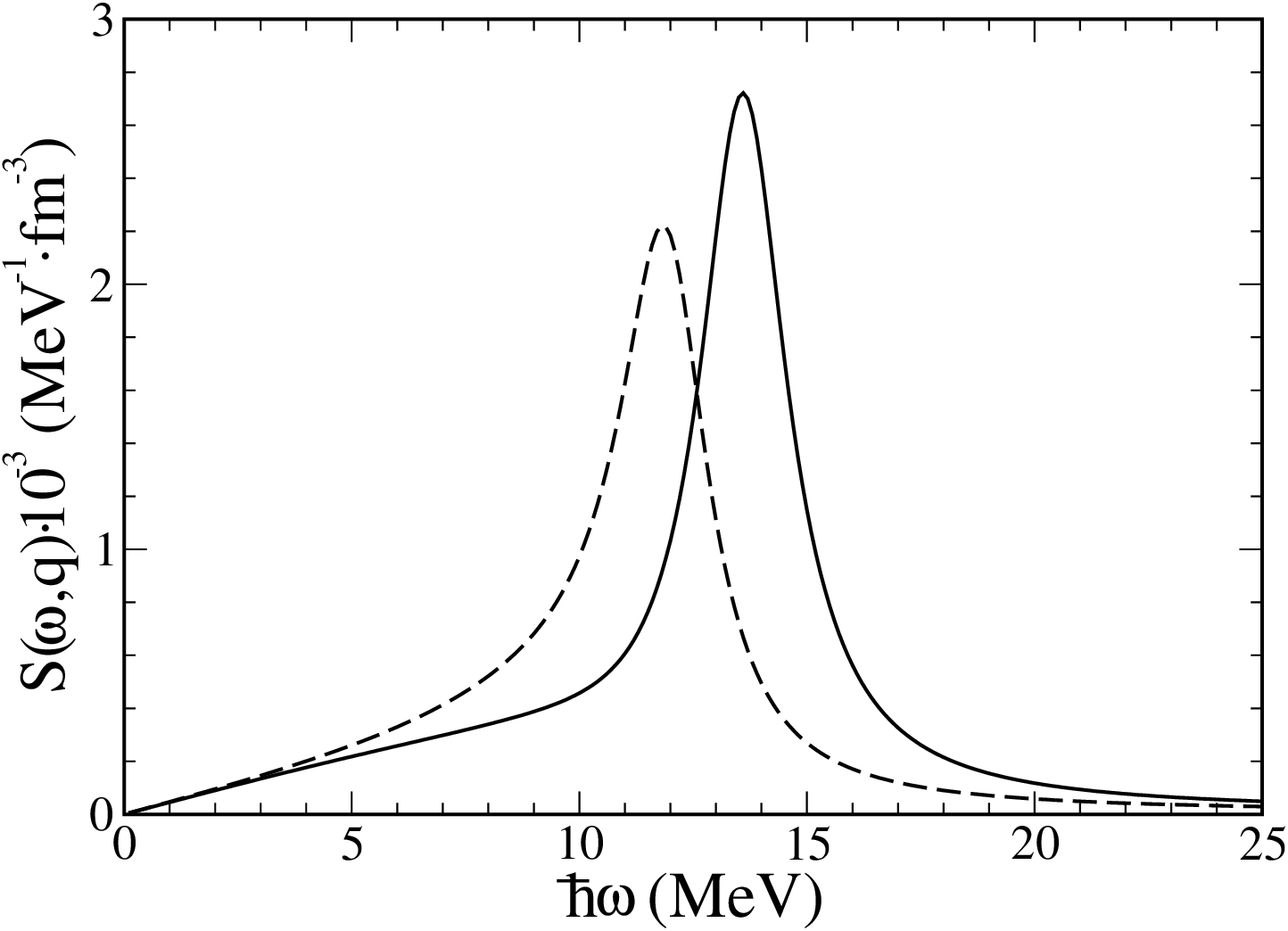}
\end{center}
\caption{Strength function $S(\omega,q)$ from Eqs. (\ref{chi2}) and (\ref{A10})
for $T=2$ MeV, $\alpha_{2}=5.4$ MeV, $F_{1}=-\ 0.3$, $\eta =-\ 0.1$, $A=208$.
Solid line for $F_{1}^{\prime }=1.1$, and dashed line for $F_{1}^{\prime}=0$.}
\label{fig3}
\end{figure}

\bigskip

Sensitivity of the strength function $S(\omega ,q)$ to the interaction
parameter $F_{1}^{\prime }$ is demonstrated in \figurename\ \ref{fig3}). The inclusion of the
nonlocal interaction $F_{1}^{\prime }\neq 0$ increases the isovector
stiffness coefficient and shifts the IVGDR to the higher energy. Moreover,
since the interaction parameter $F_{1}^{\prime }$ enters the enhancement
factor $1+\kappa _{NM}$ of Eq. (\ref{kappa1}), the photoabsorption cross
section $\sigma _{\mathrm{abs}}(\omega )$ (\ref{sigma1}) grows with $%
F_{1}^{\prime }>0$.

\begin{figure}
\begin{center}
\includegraphics*[width=9cm]{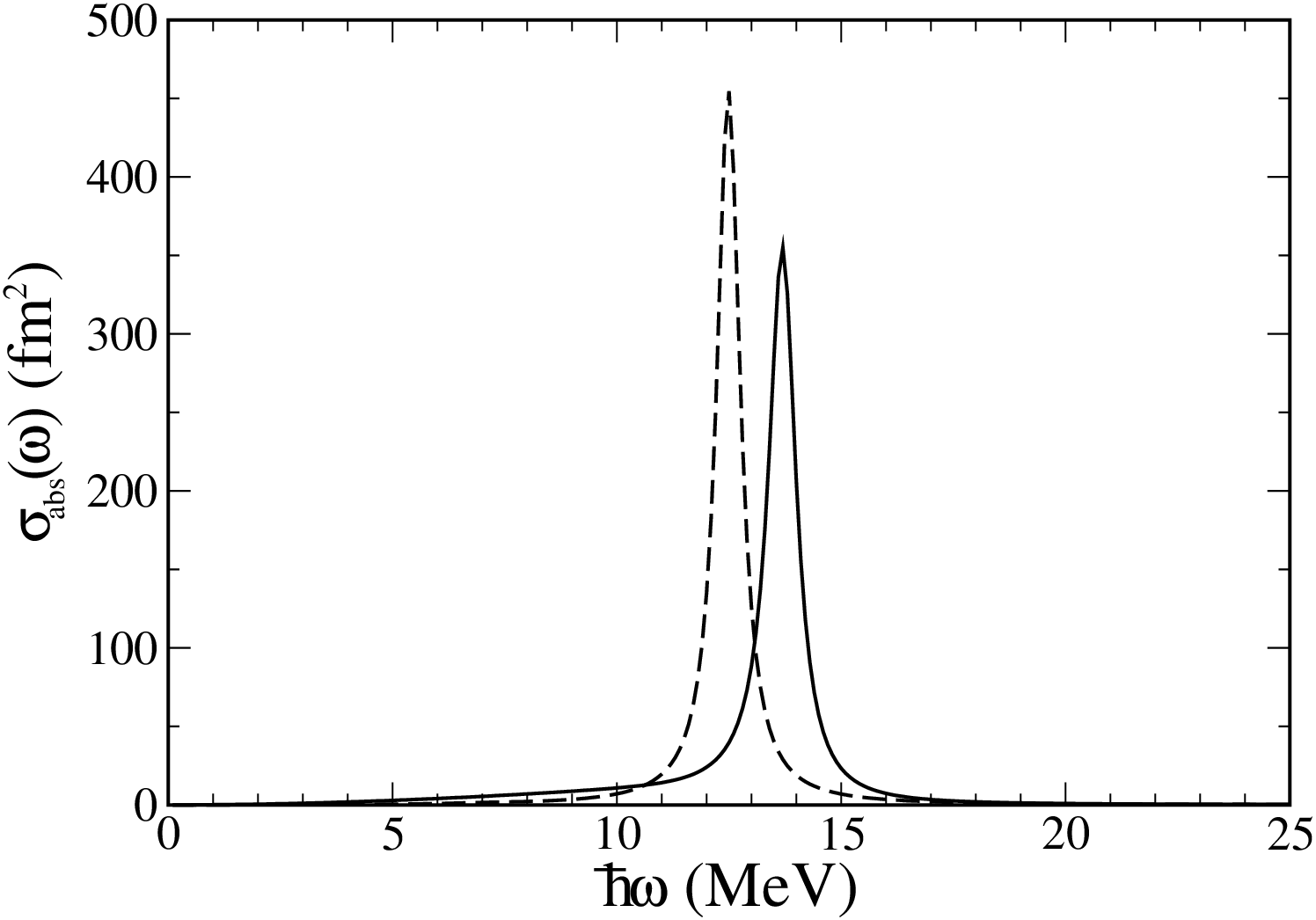}
\end{center}
\caption{Photoabsorption cross section $\sigma_{\mathrm{abs}}(\omega)$ from Eqs.
(\ref{chi2}), (\ref{sigma1}) and (\ref{A10}) for $T=2$ MeV, $A=208$, $\eta =1$,
$F_{1}=0$, $F_{1}^{\prime}=0$. Solid line for $\alpha_{2}=5.4$ MeV (zero-sound
regime) and dashed line for $\alpha_{2}=0.1$ MeV (first-sound regime).}
\label{fig4}
\end{figure}

Presence of the $1/\tau _{1}$ term in the collision integral in Eq. (\ref%
{st3}) increases the width of the IVGDR resonance and does not much affect
its energy centroid. However, this term influences strongly the zero- to
first-sound transition for the isovector mode \cite{laca99,baco99}. In
general, a decrease of the collisional relaxation time $\tau _{2}$ leads to
the fast damping of the Fermi surface distortions and thereby to the zero-
to first-sound transition. This is demonstrated in \figurename\ \ref{fig4}.

\bigskip

The solid line in \figurename\ \ref{fig4} shows the numerical result for the photoabsorption
cross section $\sigma _{\mathrm{abs}}(\omega )$ from Eq. (\ref{sigma1}) for
long relaxation time regime (zero-sound regime, $\alpha _{2}=5.4$ MeV)\ at $%
\eta =1$, i.e., $\tau _{1}=\infty $. To show the zero- to first-sound
transition, we have plotted in \figurename\ \ref{fig4} (dashed line) the cross section $%
\sigma _{\mathrm{abs}}(\omega )$ which is obtained in the first sound regime
$\omega _{R}\tau _{2}\ll 1$ at $\alpha _{2}=0.1$ MeV. This transition
happens as a shift of the resonance energy to the energy of the first sound
eigenmode, $E_{\mathrm{first}}$, given by Eq. (\ref{eres1})
\[
E_{\mathrm{first}}\approx 17.5\ \mathrm{MeV\quad for}\quad A=208.
\]%
This value of $E_{\mathrm{first}}$ significantly exceeds the \textrm{SJ}
estimate $E_{\mathrm{SJ}}\approx 14.6$ MeV for $A=208$ obtained at the
boundary condition $j_{1}^{\prime }(qR_{0})=0$\ \cite{bomo75}, see also Eqs.
(\ref{eres1}) and (\ref{esj}).

\begin{figure}
\begin{center}
\includegraphics*[width=9cm]{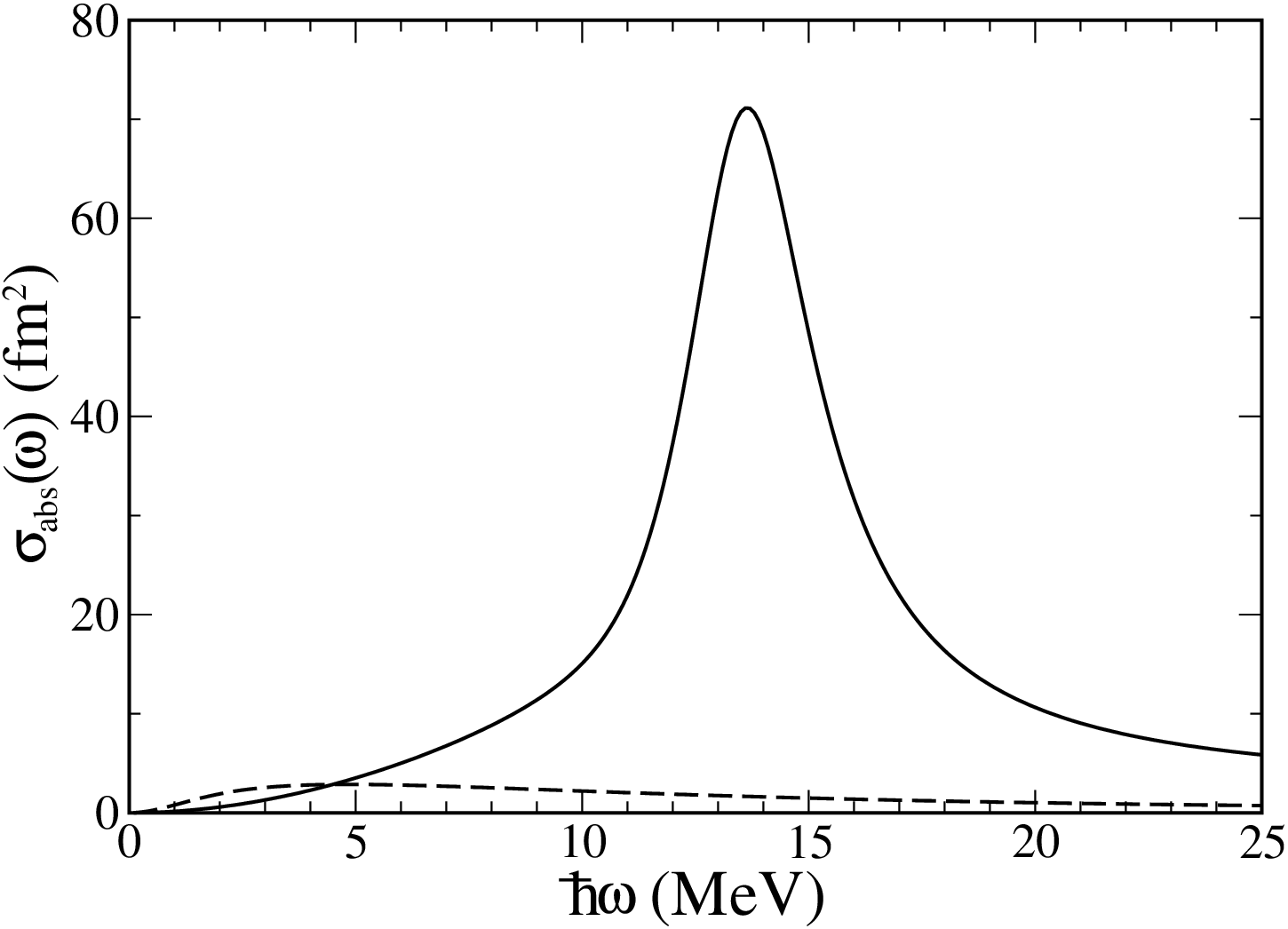}
\end{center}
\caption{The same as \figurename\ \ref{fig4}, but for $\eta =-1$.}
\label{fig5}
\end{figure}

The behavior of the photoabsorption cross section $\sigma _{\mathrm{abs}%
}(\omega )$ is essentially different for the case where the relaxation time $%
\tau _{2}$ exceeds significantly the relaxation time $\tau _{1}$. In this
case the relaxation of the relative proton-neutron motion is faster than the
zero- to first-sound transition and the first-sound peak of the IVGDR
disappears. This effect is shown in \figurename\ \ref{fig5} where the dashed line was
obtained at $\eta =-1$, i.e., $\tau _{2}=2\tau _{1}$.

\subsection{Energy-weighted sums and centroid energies}

\begin{figure}
\begin{center}
\includegraphics*[width=9cm]{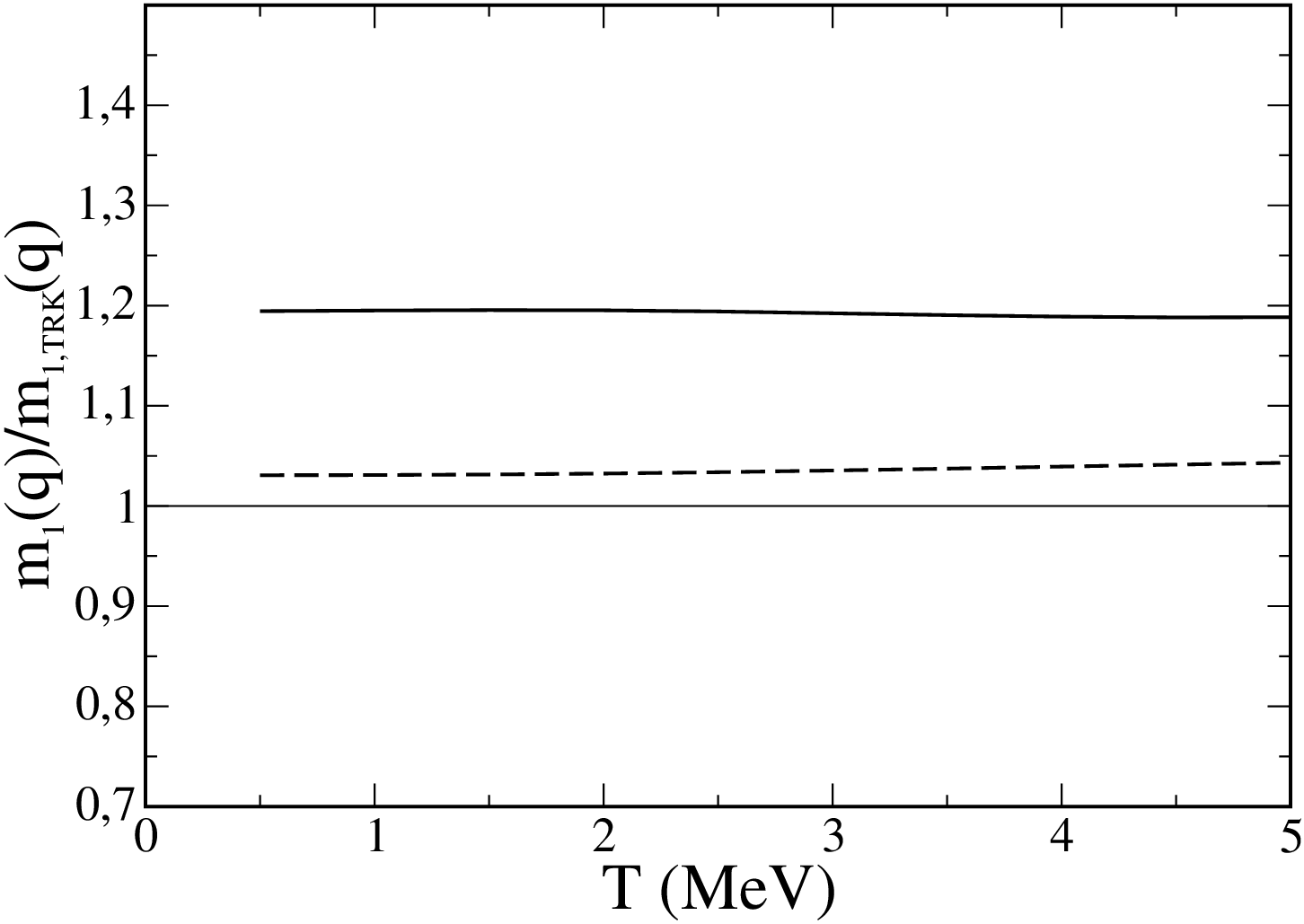}
\end{center}
\caption{Temperature dependence of the "model independent" EWS $m_{1}(q)$ obtained
from Eqs. (\ref{mk1}), (\ref{str1}) and (\ref{chi2}) normalized to the TRK sum rule
$m_{1,TRK}(q)$. The calculations performed for $A=208$ using $\alpha_{2}=5.4$ MeV,
$\eta =-\ 0.1$ and $F_{1}=-\ 0.3$. Solid line for $F_{1}^{\prime }=1.1$ and dashed
line for $F_{1}^{\prime }=0$.}
\label{fig6}
\end{figure}

We have studied the temperature behavior of \ the "model independent" EWS $%
m_{1}(q)$ and the enhancement factor $m_{1}(q)/m_{1,\mathrm{TRK}}(q)$,
where, see Eq. (\ref{mk1}),
\[
m_{1,\mathrm{TRK}}(q)=\int\limits_{0}^{\infty }d(\hbar \omega )\ \hbar
\omega \ S(\omega ,q)\quad \mathrm{for}\quad F_{1}=F_{1}^{\prime }=0.
\]%
For non-zero temperatures and in presence of the relaxation, the
energy-weighted sum $m_{1}(q)$ has been evaluated using the definition (\ref%
{mk1}) and the response function $\chi (\omega ,q)$ from Eq. (\ref{chi2}).
In \figurename\ \ref{fig6}, we have plotted the ratio $m_{1}(q)/m_{1,\mathrm{TRK}}(q)$ as a
function of temperature $T$.

\bigskip

As can be seen from \figurename\ \ref{fig6}, the enhancement factor is only slightly
sensitive to the temperature variation. For the nucleus $^{208}Pb$, we have
from \figurename\ \ref{fig6} the following estimate
\[
m_{1}(q)/m_{1,\mathrm{TRK}}(q)\approx 1.2.
\]

\begin{figure}
\begin{center}
\includegraphics*[width=9cm]{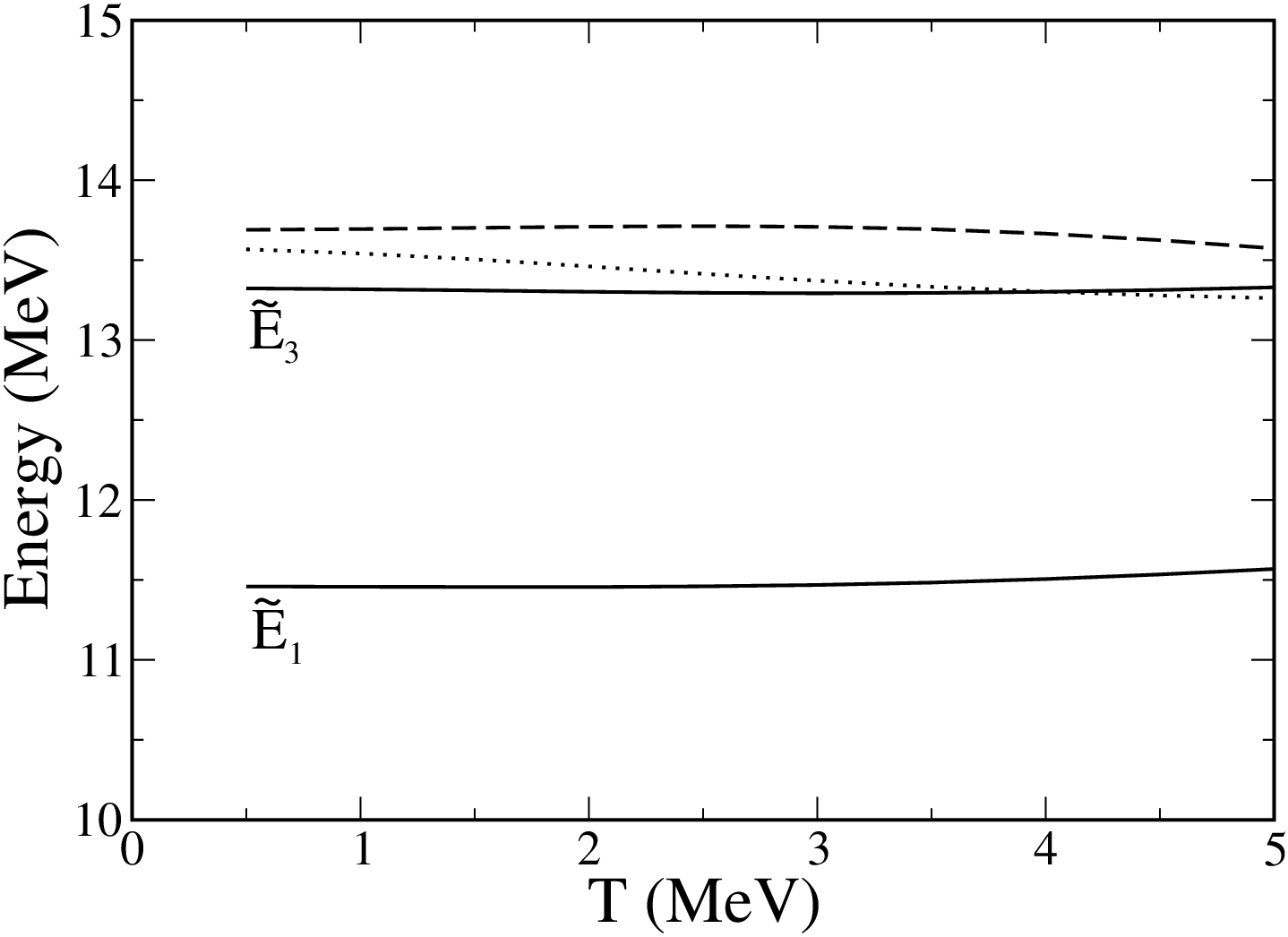}
\end{center}
\caption{Temperature dependence of the centroid energies
$\widetilde{E}_{1}=\sqrt{m_{1}/m_{-1}}$ (solid line) and
$\widetilde{E}_{3}=\sqrt{m_{3}/m_{1}}$ (solid line) obtained
from Eq. (\ref{mk1}), and the IVGDR eigenenergy $E_{R}$
(dashed line) obtained from Eqs. (\ref{disp1}) and (\ref{seceq}). The
calculations were performed for $A=208$ using $\alpha_{2}=5.4$ MeV,
$\eta =2/3$, $m^{\ast }/m=0.9$ and $F_{1}^{\prime }=1.1$. Dotted line shows
the classical energy $E_{\mathrm{macr}}$ from Eq. (\ref{eclas1}).}
\label{fig7}
\end{figure}

In \figurename\ \ref{fig7}, we compare the centroid energies $\widetilde{E}_{1}=\sqrt{%
m_{1}/m_{-1}}$ and $\widetilde{E}_{3}=\sqrt{m_{3}/m_{1}}$ evaluated using
Eq. (\ref{mk1}), and the IVGDR\ eigenenergy $E_{R}$ obtained from Eqs. (\ref%
{disp1}) and (\ref{seceq}).

\bigskip

The significant upward shift of energy $\widetilde{E}_{3}$ curve with
respect to $\widetilde{E}_{1}$ is due to the Fermi surface distortion
effect. The cubic sum $m_{3}$ which enters the expression for $\widetilde{E}%
_{3}$ is associated with the scaling transformation and the quadrupole
distortion of the Fermi surface \cite{kosh04}. On the other hand, the
distortion of the Fermi surface causes an additional contribution to the
stiffness coefficient (see also Eqs. (\ref{c2}) and (\ref{cq2})) providing a
growth of the centroid energy $\widetilde{E}_{3}$. In contrast to $%
\widetilde{E}_{3}$, the centroid energy $\widetilde{E}_{1}$ is derived by
the inverse sum $m_{-1}$, where the contribution from the Fermi-surface
distortion effects is negligible, see also Eq. (\ref{cq1}). Note that the
presence of the Fermi-surface distortion effects increases also the
resonance energy $E_{R}$ because the dispersion equation (\ref{disp1})
includes\ all multipolarities of the Fermi surface distortion. A small
decrease of the energy $E_{R}$ with growing temperature $T$ in \figurename\ \ref{fig7} is due
to the fact that the Fermi-surface distortion effects become weaker for
higher temperatures. In contrast to $\widetilde{E}_{3}$, which is nearly
temperature independent, the centroid energy $\widetilde{E}_{1}$ increases
slightly with $T$. That is because a growth of resonance width with
temperature leads to a decrease of the inverse sum $m_{-1}$ in agreement
with its definition\ of Eq. (\ref{mk1}).

In \figurename\ \ref{fig7}, the dotted line shows the behavior of the eigenenergy $E_{\mathrm{%
macr}}$ for the macroscopic eigenvibrations given by Eq. (\ref{eclas1}). To
provide the convergency of the integral in Eq. (\ref{mk1}) for the inverse
sum $m_{-3}$ in Eq. (\ref{eclas1}), we have introduced the cut off parameter
$E_{\mathrm{cut}}$ in Eq. (\ref{mk1}). Namely, we have used
\begin{equation}
m_{-3}(q)=\int\limits_{E_{\mathrm{cut}}}^{\infty }d(\hbar \omega )\ (\hbar
\omega )^{-3}\ S(\omega ,q).  \label{mk7}
\end{equation}%
The cut off energy $E_{\mathrm{cut}}$ was derived from the requirement that
the Landau-damping region $s<1$ should be removed from the EWS $m_{k}$ since
$m_{k}$ must be related to a given sound eigenmode only. Note that the
energy interval $\left[ E_{\mathrm{cut}},\infty \right] $ must be large
enough to provide a reasonable exhaustion ($\gtrsim 90\%$) of sum rule
(\ref{mk6}). A good agreement of $E_{\mathrm{macr}}$ with the IVGDR\ eigenenergy
allows one to conclude that the cut off procedure in Eq. (\ref{mk7}) can
also be used for the consistent evaluation of the mass coefficient $B_{Q}$
of Eq. (\ref{b1}).

Finally, it should be noted that there is the limiting temperature
$T_{\mathrm{lim}}$ for the IVGDR existence. The limiting temperature
can be deduced from the decomposition of the
$\widetilde{\chi}_{\mathrm{in}}(\omega,q)$ of Eq. (\ref{A6}) in
powers of $1/s$ for the first sound limit (high temperature regime)
at $|s|\ \gg 1$. The presence of the finite relaxation time
$\tau_{1}$ caused by the collisions for the counterstreaming neutron
and proton flows leads to the additional temperature dependence of
the eigenfrequency which is specific for the isovector mode only.
The eigenfrequency of the isovector first sound decreases with
temperature and disappears at the limiting temperature
$T_{\mathrm{lim}}$. The magnitude of the limiting temperature
$T_{\mathrm{lim}}$ for the IVGDR  depends significantly on the
relaxation parameter $\eta$. For heavy nuclei, the numerical
estimate \cite{laca99} provides the value of $T_{\mathrm{lim}}\simeq
7$ MeV.

\subsection{Damping and spreading width of IVGDR}

The EWS $m_{k}(q)$ can be used to analyze the spreading of the strength
function. We will use the following definition of the spreading width \cite%
{list89,hasa97}
\begin{equation}
\gamma (q)=\frac{m_{1}(q)}{m_{0}(q)}-\frac{m_{0}(q)}{m_{-1}(q)}.
\label{gam1}
\end{equation}%
In the case of a Lorentzian shape of the photoabsorption cross section
\begin{equation}
\sigma _{\mathrm{abs}}(\omega )=\frac{\sigma _{0\ }E^{2}\Gamma ^{2}}{%
(E-E_{0})^{2}+E^{2}\Gamma ^{2}},\qquad E=\hbar \omega ,  \label{sigma2}
\end{equation}%
one can find from Eqs. (\ref{sigma1}), (\ref{mk1}) and (\ref{gam1}) the
relationship between the quantities $\widetilde{E}_{1}$ and $\gamma (q)$ and
the resonance characteristics $E_{0}$ and $\Gamma $
\begin{equation}
E_{0}=\widetilde{E}_{1},\quad \gamma (q)=\frac{2}{\pi }\Gamma \left(
1+O(\Gamma /E_{0})\right) \quad \textrm{for}\quad \Gamma /E_{0}\ll 1.
\label{expr1}
\end{equation}

In the case of small damped collective vibrations $(\Gamma \ll E_{0})$, the
resonance collisional width $\Gamma$ can also be evaluated from the
dispersion equation (\ref{disp1}). The solution of this equation
\begin{equation}
\omega =\omega _{R}+i\omega _{I},
\label{disp3}
\end{equation}
defines the energy of the giant resonance $E_{R}=\hbar \omega _{R}$ and its
collisional width $\Gamma_{\mathrm{col}}=-2\hbar \omega _{I}$. In the
application of the kinetic approach, in particular, of the collision
integral $\mathrm{\delta St}[f]$ to\ the finite nucleus, the difficulty is
the derivation of the $NN$ cross sections $\sigma_{\mathrm{av}}$ and
$\sigma_{-}$ in Eq. (\ref{alpha1}) which become $\vec{r}$-dependent in nuclear
interior. The above mentioned (see comment to Eq. (\ref{alpha1}))
vacuum and in-medium values of the $NN$ cross sections and
thereby relaxation parameter $\alpha _{2}$ in Eqs. (\ref{tau1}) and
(\ref{alpha1}) give the lower and upper theoretical limits. The $\vec{r}$-dependence
of the relaxation parameter $\alpha_{2}$ can
be taken into account phenomenologically by introduce of the effective
relaxation parameter $\alpha_{\mathrm{eff}}$ as following
\begin{equation}
\frac{1}{\alpha_{\mathrm{eff}}}=\frac{\displaystyle\int d\vec{r}
\rho_{\mathrm{eq}}(\vec{r})/\alpha_{2}(\vec{r})}
{\displaystyle\int d\vec{r}\rho_{\mathrm{eq}}(\vec{r})},
\label{alphaeff}
\end{equation}
where $\alpha(\vec{r})$ is parametrized by
\[
\alpha_{2}(\vec{r})=\alpha_{\mathrm{vac}}
+\frac{\rho_{\mathrm{eq}}(\vec{r})}{\rho_{\mathrm{eq}}(0)}
\left(\alpha_{\mathrm{bulk}}-\alpha_{\mathrm{vac}}\right),
\]
with $\alpha_{\mathrm{vac}}=2.3$, MeV and $\alpha_{\mathrm{bulk}}=5.4$, MeV,
see comments after Eq. (\ref{alpha1}). The equilibrium particle density
$\rho_{\mathrm{eq}}(\vec{r})$ is derived as
\[
\rho_{\mathrm{eq}}(\vec{r})=\rho_{0}\left/\left[ 1+\exp \frac{r-R_{0}}{a}\right]\right.
\]
where $\rho_{0}=(4\pi r_{0}^{3}/3)^{-1}$ and $a=0.6$ fm. Below we
will replace relaxation parameter $\alpha_{2}$ in Eq. (\ref{tau1}) by the
effective one $\alpha_{\mathrm{eff}}$.

Besides the collisional width $\Gamma_{\mathrm{col}}$, the experimentally
observable width of the IVGDR includes the fragmentation width. Within the
semiclassical kinetic theory, this mechanism of resonance spreading can be
considered as an additional dissipation due to the single particle
scattering on the moving surface of the nucleus (one-body dissipation
\cite{mysw77,blbo78,siko78}). Instead of $\tau_{2}$, we will use the effective
relaxation time $\tau_{\mathrm{eff}}$ which contains the contribution from
both two-body and one-body dissipations. Namely, see also Refs.
\cite{laca99,kopl96},
\begin{equation}
\frac{1}{\tau_{\mathrm{eff}}}=\frac{1}{\tau _{2}}+\frac{1}{\tau _{\mathrm{%
wall}}}.  \label{taueff}
\end{equation}%
We will use the one-body relaxation time $\tau _{\mathrm{wall}}$ in the
following form \cite{kopl96}
\begin{equation}
\tau _{\mathrm{wall}}=\frac{2R_{0}}{\bar{v}}\xi, \quad \bar{v}=\frac{3v_{F}}{4}
\left[1+\frac{\pi^{2}}{6}\left(\frac{T}{\epsilon_{F}}\right)^{2}\right].
\label{tauwall}
\end{equation}
The parameter $\xi$ in Eq. (\ref{tauwall}) depends on the model of the
one-body dissipation \cite{mysw77,blbo78,siko78,nisi80}. We consider $\xi$
as a free parameter which is determined from a fit of the total IVGDR width
$\Gamma$ to the experimental data at zero temperature $T=0$.

\begin{figure}
\begin{center}
\includegraphics*[width=9cm]{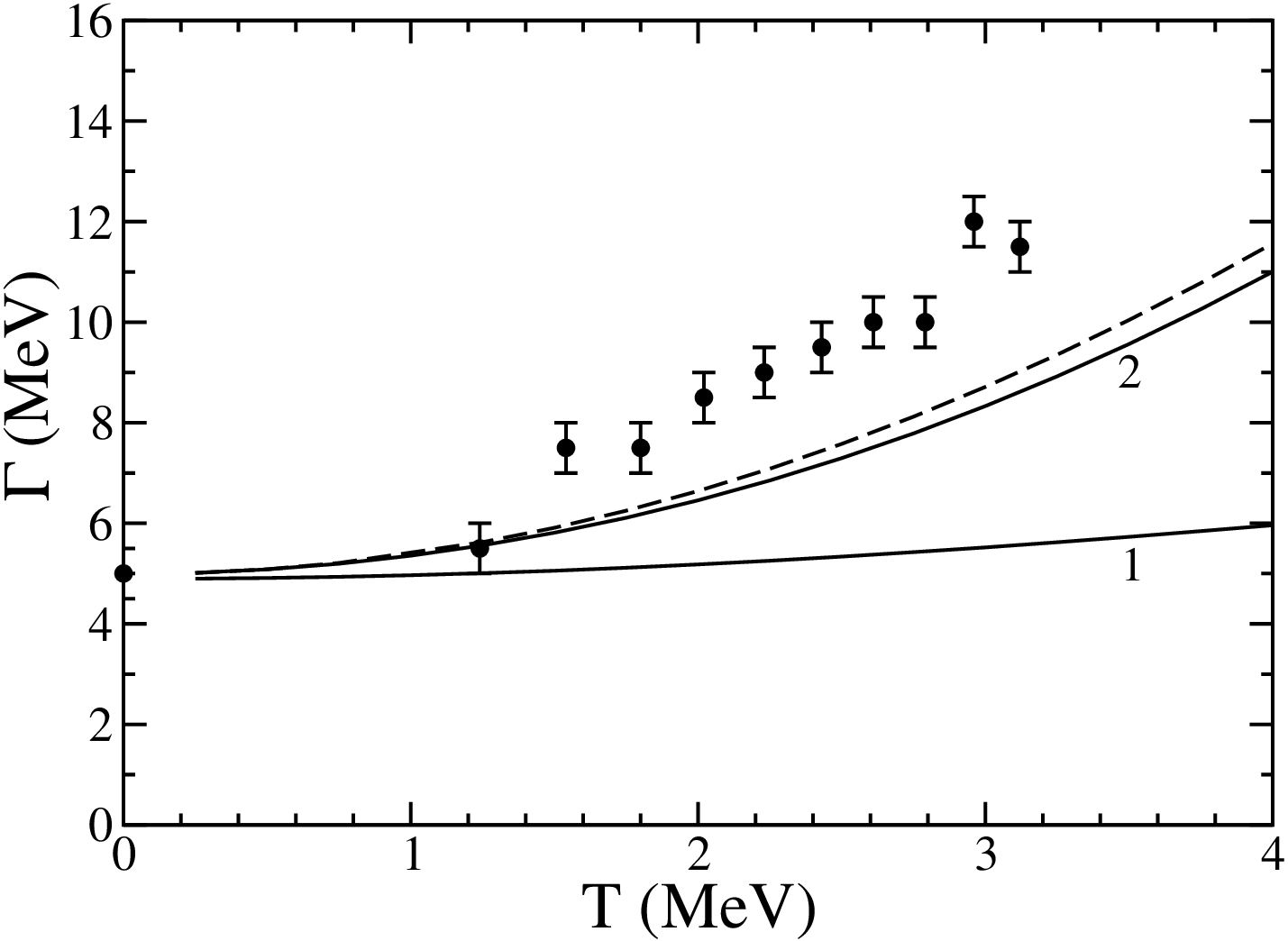}
\end{center}
\caption{Temperature dependence of the IVGDR width $\Gamma$: the solid line
1 was obtained from the EWS definition given by Eqs. (\ref{gam1}) and (\ref{expr1});
the solid line 2 and the dashed line were obtained from the dispersion equation
(\ref{disp1}). The calculations were performed for the nucleus $^{120}Sn$
using $\alpha_{\textrm{eff}}$ from Eq. (\ref{alphaeff}) and $\eta =-0.1$.
Solid line 2 for $F_{1}=-\ 0.3$, $F_{1}^{\prime }=1.1$ and dashed line for
$F_{1}=F_{1}^{\prime}=0$.}
\label{fig8}
\end{figure}

In \figurename\ \ref{fig8} we have plotted the temperature dependence of the
width $\Gamma$ derived from the EWS by use of Eq. (\ref{expr1}) (solid line 1)
and from the dispersion relation applying Eqs. (\ref{disp1}), (\ref{disp3}),
(\ref{alphaeff}) and (\ref{taueff}) (solid line 2). The numerical calculations
were performed for the nucleus $^{120}Sn$ where the experimental
data are known for a wide range of temperatures $T=0\div 3$ MeV
\cite{rama96a,rama96b,baum98}.

We point out that considering the experimental data we assume that
the nuclear excitation energy $E^{\ast}$ is related to
the nuclear themperature $T$ by the Fermi-gas formula $T=\sqrt{E^{\ast}/a}$
where $a$ is level density parameter. This fact leads to some uncertainty
because the derivation of parameter $a$ is model dependent. However, within
the kinetic theory, the use of the macrocanonical ensemble, and thereby
temperature $T$, is preferable because the ensemble smearing is assumed at
the derivation of basic collisional kinetic equation.

\bigskip

In \figurename\ \ref{fig8}, the total resonance width $\Gamma$ grows with temperature
mainly due to the temperature dependence of the collisional relaxation time $\tau_{2}$,
see Eq. (\ref{tau1}). As can be seen from Eq. (\ref{tauwall}), the temperature dependence
of the one-body dissipation is too weak to be responsible for the fast increase of the
IVGDR width. The result in \figurename\ \ref{fig8} shows also that the collisional width
does not give a good description of the observable growing of the IVGDR width with $T$.
An additional increase of the IVGDR width with temperature can be achieved taking into
consideration the coupling of the IVGDR to thermal shape fluctuations \cite{orbo96}
which do not present in our approach. Note that a decrease of the in-medium $NN$ cross
section leads to an increase of the temperature dependence of the IVGDR width
\cite{diko99} also.

For low temperatures, the total resonance width $\Gamma $ derived by the EWS
(solid line 1 in \figurename\ \ref{fig8}) is close to the one obtained from the dispersion
relation (solid line 2 in \figurename\ \ref{fig8}). That is due to the fact that the
evaluated photoabsorption cross section approaches to the Lorentzian shape
in this case. For higher temperatures, the assumption $\Gamma /E_{0}\ll 1$
in Eq. (\ref{expr1}) is not fulfilled and the expressions (\ref{gam1}) and (%
\ref{expr1}) for the derivation of $\Gamma $ can not be correctly used.

\section{Summary and Conclusions}

A goal of this work is the derivation of the macroscopic characteristics of
isovector modes in a two-component Fermi-liquid drop from our knowledge of
the nuclear IVGDR. Starting from the collisional kinetic equation (\ref{eq1}%
), we have derived the strength function and the energy weighted sums $m_{k}$
for the isovector excitations in the heated nuclear matter and the finite
nuclei. An important ingredient of our consideration is the inclusion of the
velocity dependent $NN$-interaction for both isovector and isoscalar
channels. Our consideration is valid for an arbitrary collision parameter $%
\alpha _{l}$ in Eq. (\ref{tau1}) and can be used, particularly, for the
transition region from the zero sound- to first sound (hydrodynamic) regime
in nuclear Fermi-liquid.

We have adopted a Fermi liquid drop model with two essential features: (i)
The linearized kinetic equation is applied to the nuclear interior, where
the relatively small oscillations of the particle density take place; (ii)
The dynamics in the surface layer of the nucleus is described by means of
the macroscopic boundary condition which is taken as a condition for the
balance of the forces on the free nuclear surface.

Due to the consistent solution of the dispersion relation (\ref{disp1}) and
secular equation (\ref{seceq}), our model provides a satisfactory
description of the $A$-dependence of the eigenenergy of the IVGDR, see \figurename\ \ref{fig1}.
In contrast to earlier consideration of Ref. \cite{stri83} performed
within the scaling approximation, we take into account all multipolarity $l$
of the Fermi surface distortion. Moreover, the value of Landau's interaction
parameter $F_{1}^{\prime }\approx 1.1$ has been derived from a fit of the
evaluated EWS enhancement factor to the experimental data. Note that we do
not use a concept of the effective $NN$-interaction within the surface layer
of the nucleus applying instead the relevant boundary condition and avoiding
thereby the uncertainty in the derivation of effective interaction in the
region of strong particle density inhomogeneity.

The present study has shown the following:

\begin{enumerate}
\item The Landau damping occurs in the isovector strength function $S(\omega,q)$ 
at low temperatures as a wide bump on the left side of the narrow sound
peak (see \figurename\ \ref{fig2}). For high temperature, the overlapping of 
both the sound peak and the Landau damping bump leads to the asymmetry of the 
IVGDR strength function $S(\omega ,q)$. A similar asymmetry of the IVGDR strength
function is also observed experimentally at non-zero temperatures \cite{baum98}.

\item The isovector EWS shows only minor temperature dependence. In particular, 
the "model \ independent" \ \ EWS \ \ $m_{1}(q)$ \ \ and  \ the corresponding \
enhancement \ factor \\ $m_{1}(q)/m_{1,TRK}(q)$ are practically constant in the
interval of temperature $T=0\,\div \,5$ MeV, see \figurename\ \ref{fig6}.

\item The "model independent" EWS $m_{1}(q)$ and the enhancement factor 
$m_{1}(q)/m_{1,TRK}(q)$ are slightly sensitive to the relaxation (damping)
processes. Note that the corresponding problem can not be accurately
considered within the standard quantum mechanics because of non-Hermite
Hamiltonian in this case.

\item The inclusion of the nonlocal interaction $F_{1}^{\prime }\neq 0$
increases the isovector stiffness coefficient and shifts the IVGDR energy to
the higher values. The lowest order EWS $m_{-1}(q),\ m_{1}(q)$ and $m_{3}(q)$
derive the adiabatic, $\widetilde{E}_{1}$, and scaled, $\widetilde{E}_{3}$,
energy centroids. The centroid energy $\widetilde{E}_{1}$ is close to the
classical result of the Steinwedel-Jensen model and differs from the\ first
sound limit by the enhancement factor $1+\kappa _{NM}$, see Eq. (\ref{even1}%
). The Fermi distortion effects do not contribute into the centroid energy $%
\widetilde{E}_{1}$. In contrast, the scaling energy $\widetilde{E}_{3}$ is
associated with the quadrupole distortion of the Fermi surface and exceeds
significantly the centroid energy $\widetilde{E}_{1}$.

\item The often used classical derivation of the IVGDR eigenenergy $E_{\mathrm{%
macr}}$ Eq. (\ref{eclas1}) through the isovector mass coefficient $B_{Q}$
has to be revised in the case of nuclear Fermi liquid. That is because the
inverse sum $m_{-3}$, which enters\ the mass coefficient $B_{Q}$,\ can not
be directly evaluated within the kinetic theory due to the Landau-damping
region at $\omega \sim 0$. To provide the convergency of the sum $m_{-3}$,
we have proposed the cut-off procedure introducing an appropriate cut off
parameter $E_{\mathrm{cut}}$ into the definition of $m_{-3}$ in Eq. (\ref%
{mk1}). Due to this procedure we have achieved a good agreement of $E_{%
\mathrm{macr}}$ with the adiabatic centroid energy $\widetilde{E}_{1}$. In
general, such kind of cut-off procedure can also be used for the evaluation
of the transport coefficients within the Fermi-liquid theory.

\item The inclusion into the collision integral of the relaxation of $l=1$
component (term $\sim 1/\tau _{1}$ in Eq. (\ref{st3})) influences strongly
the zero- to first-sound transition for the isovector mode. In particular,
in the case of $\tau _{2}>\tau _{1}$, the relaxation of the relative
proton-neutron motion is too fast and the short-relaxation limit $\alpha
_{2}\rightarrow 0$ does not provide the zero- to first-sound transition (see
disappearance of the first-sound peak in \figurename\ \ref{fig5}). A similar phenomenon was
earlier discussed in Refs. \cite{laca99,baco99} where the velocity
independent Landau's interaction with $F_{1}^{\prime }=0$ has been used.

\item Our analysis of the IVGDR width performed within the kinetic theory
shows that the collisional and one-body damping does not reproduce the fast
increase of the IVGDR width with temperature. The additional mechanisms of
damping, e.g., the coupling of the IVGDR to thermal shape fluctuations \cite%
{orbo96}, have to be involved to improve the agreement of the temperature
dependence of the IVGDR width with the experimental data.
\end{enumerate}

Finally, we would like to note that the semiclassical kinetic approach, used
in this article, is highly convenient for a study of the averaged properties
of the nuclear dynamics. This approach provides an information on the
macroscopic characteristics without a detailed knowledge of the wave
function of the nucleus. An important advantage of the kinetic theory is
that the temperature and the relaxation effects enter the equation of motion
directly. Many results can be here presented in a clear and transparent
form. Here, the claim is to describe the general features of collective
excitations, such as the $A$-dependence of the IVGDR energy, in a systematic
way ignoring many quantum effects, e.g., the shell structure effects. In
particular, the kinetic approach allows us to compare the results of
standard liquid drop model \cite{bomo75} with the Fermi liquid drop one
where the dynamic distortions of Fermi surface are taken into consideration.

\bigskip

\bigskip \setcounter{equation}{0}\renewcommand{\theequation}
{A \arabic{equation}} \noindent\textbf{Appendix A: Internal response function}
\smallskip

Using Eq. (\ref{eq3}), we will evaluate the averaged amplitude $\widetilde{\nu}_{0}$.
Taking into account Eq. (\ref{int2}), we will reduce Eq. (\ref{eq3}) to the following
coupled equations
\begin{equation}
\nu_{0}(p)+F_{0}^{\prime}Q_{00}(z)\widetilde{\nu}_{0}
+\frac{F_{1}^{\prime}}{3}Q_{10}(z)\widetilde{\nu}_{1}
-\lambda_{0}Q_{00}(z)
=i\gamma\nu_{0}(p)\frac{1}{z}\left[1-Q_{00}(z)\right]
-i\eta\gamma\nu_{1}(p)Q_{00}(z),
\label{A1}
\end{equation}
\begin{equation}
\nu_{1}(p)+3F_{0}^{\prime}Q_{10}(z)\widetilde{\nu}_{0}
+F_{1}^{\prime}Q_{11}(z)\widetilde{\nu}_{1}
-3\lambda_{0}Q_{10}(z)
=-3i\gamma\nu_{0}(p)Q_{00}(z)-3i\eta\gamma\nu_{1}(p)Q_{10}(z),
\label{A2}
\end{equation}
where we have used $Q_{10}(z)=z\ Q_{00}(z)$. Solving Eqs. (\ref{A1}) and (\ref{A2}) 
we obtain

\begin{equation}
\nu_{0}(p)=\frac{z\ \chi _{0}(z)}{sD(z)+i\gamma \chi _{0}(z)\ }\lambda
_{0}-F_{0}^{\prime }\frac{z\ \chi _{0}(z)}{sD(z)+i\gamma \chi_{0}(z)}
\widetilde{\nu }_{0}
-\frac{F_{1}^{\prime }}{3}\frac{z\chi _{0}(z)(z-i\eta \ \gamma )}
{sD(z)+i\gamma \chi _{0}(z)}\widetilde{\nu }_{1}
\label{A3}
\end{equation}
and
\[
\nu _{1}(p)=\frac{3z\ s\ \chi _{0}(z)}{sD(z)+i\gamma \chi _{0}(z)\ }\lambda
_{0}-F_{0}^{\prime }\frac{3z\ s\ \chi _{0}(z)}{sD(z)+i\gamma \chi _{0}(z)}%
\widetilde{\nu }_{0}
\]
\begin{equation}
-F_{1}^{\prime }\frac{\left[ z^{2}\chi _{0}(z)+1/3\right] \ \left[
sD(z)+i\gamma \chi _{0}(z)\right] -i\gamma \ z\ (z-i\eta \ \gamma )\chi
_{0}^{2}(z)}{D(z)\left[ sD(z)+i\gamma \chi _{0}(z)\right] }\widetilde{\nu }%
_{1},  \label{A4}
\end{equation}%
where $\chi _{0}(z)=Q_{00}(z),$

\[
D(z)=1+3i\eta \gamma z\chi _{0}(z)
\]%
and the relation $Q_{11}(z)=z^{2}\chi _{0}(z)+1/3$ has been used.
Multiplying both Eqs. (\ref{A3}) and (\ref{A4}) by $-\left[ g/(2\pi \hbar
)^{3}N_{T}\right] \left( \partial f_{\mathrm{eq}}/\partial \epsilon
_{p}\right) $, integrating over $\vec{p}$ and using Eqs. (\ref{nT}) and (\ref%
{nul1}), we obtain two closed equations for amplitudes $\widetilde{\nu }_{0}$
and $\widetilde{\nu }_{1}$. Solving then these equations with respect to $%
\widetilde{\nu }_{0}$, we obtain

\begin{equation}
\widetilde{\nu }_{0}=\frac{\widetilde{\chi }_{\mathrm{in}}(\omega ,q)}{%
1+F_{0}^{\prime }\ \widetilde{\chi }_{\mathrm{in}}(\omega )}\lambda _{0}.
\label{A5}
\end{equation}%
Here $\widetilde{\chi }_{\mathrm{in}}(\omega ,q)$ is the internal response
function
\begin{equation}
\widetilde{\chi }_{\mathrm{in}}(\omega ,q)=\widetilde{\chi }_{\mathrm{in}%
,0}(\omega ,q)-F_{1}^{\prime }\frac{\widetilde{\chi }_{s}^{(1)}(\omega ,q)\
\widetilde{\chi }_{\eta }^{(1)}(\omega ,q)}{1+F_{1}^{\prime }\widetilde{\chi
}^{(2)}(\omega ,q)},  \label{A6}
\end{equation}%
where
\begin{equation}
\widetilde{\chi }_{\mathrm{in},0}(\omega ,q)=-\frac{1}{N_{T}}\int \,{\frac{gd%
\vec{p}}{(2\pi \hbar )^{3}}}\,{\frac{\partial f_{\mathrm{eq}}(\epsilon _{p})%
}{\partial {\epsilon _{p}}}}\frac{z\ \chi _{0}(z)}{sD(z)+i\gamma \chi _{0}(z)%
},  \label{A7}
\end{equation}

\begin{equation}
\widetilde{\chi }_{s}^{(1)}(\omega ,q)=-\frac{1}{N_{T}}\int \,{\frac{gd\vec{p%
}}{(2\pi \hbar )^{3}}\frac{\partial f_{\mathrm{eq}}(\epsilon _{p})}{\partial
{\epsilon _{p}}}}\,\frac{z\ s\ \chi _{0}(z)}{sD(z)+i\gamma \chi _{0}(z)},
\label{A8}
\end{equation}
\begin{equation}
\widetilde{\chi }_{\eta }^{(1)}(\omega ,q)=-\frac{1}{N_{T}}\int \,{\frac{gd%
\vec{p}}{(2\pi \hbar )^{3}}\frac{\partial f_{\mathrm{eq}}(\epsilon _{p})}{%
\partial {\epsilon _{p}}}}\,\frac{z\ (z-i\eta \ \gamma )\ \chi _{0}(z)}{%
sD(z)+i\gamma \chi _{0}(z)}
\label{A9}
\end{equation}
and
\begin{eqnarray}
&&\widetilde{\chi}^{(2)}(\omega,q)=-\frac{1}{N_{T}} \int \,{\frac{gd\vec{p}}
{(2\pi\hbar)^{3}}\frac{\partial f_{\mathrm{eq}}(\epsilon_{p})}{\partial\epsilon_{p}}}
\nonumber \\
&&\times\frac{\left[z^{2}\chi_{0}(z)+1/3\right]\ \left[sD(z)+i\gamma\chi_{0}(z)\right]
-i\gamma\ z(z-i\eta\ \gamma)\chi_{0}^{2}(z)}{D(z)\left[sD(z)+i\gamma\chi_{0}(z)\right]}.
\label{A10}
\end{eqnarray}

\bigskip \setcounter{equation}{0}\renewcommand{\theequation}{B
\arabic{equation}} \noindent\textbf{Appendix B: Boundary condition}
\smallskip

In this appendix we are going to determine the boundary condition from the
balance of the forces on the free nuclear surface given by Eq. (\ref{B1}).
The internal force $\vec{F}$ in Eq. (\ref{B1}) is related to the momentum
flux tensor $\Pi_{\alpha\beta}$ in the nuclear interior
\begin{equation}
F_{\alpha }=n_{\beta }\Pi _{\alpha \beta }.  \label{B2}
\end{equation}
The momentum flux tensor $\Pi _{\alpha \beta }$ can be evaluated using the
basic kinetic equation (\ref{eq1}). Taking the $1$-st $\vec{p}$-moment of
Eq. (\ref{eq1}) one can obtain the following expression for the momentum
flux tensor $\Pi _{\alpha \beta }$ \cite{koma93,kosh04}

\begin{equation}
\Pi _{\alpha \beta }=\delta P\ \delta _{\alpha \beta }+\delta \sigma
_{\alpha \beta },  \label{B3}
\end{equation}
where $\delta P$ is the pressure caused by the isovector sound wave

\begin{equation}
\delta P=\frac{1}{3m}\int \frac{gd\vec{p}}{(2\pi \hbar )^{3}}p^{2}\ \delta
f_{\_}(\vec{r},\vec{p};t)+\frac{F_{0}^{\prime }}{N_{F}}\overline{\rho }_{%
\mathrm{eq}}\ \delta \rho _{\_}(\vec{r},t)=C_{\mathrm{sym}}\ \delta \rho
_{\_}(\vec{r},t)  \label{B4}
\end{equation}%
and $\delta \sigma _{\alpha \beta }$ is the pressure tensor due to the Fermi
surface distortion effect
\begin{eqnarray}
\delta\sigma_{\alpha\beta}&=&\frac{1}{3m}\int\frac{gd\vec{p}}{(2\pi\hbar)^3}
(3p_\alpha p_\beta -p^2\delta_{\alpha\beta})\ \delta f(\vec{r},\vec{p};t)
\nonumber \\
&=&\mu_F \left(\nabla_\alpha \chi_\beta +\nabla_\beta \chi_\alpha -
\frac{2}{3}\vec{\nabla}\cdot\vec{\chi}\, \delta_{\alpha\beta}\right).
\label{B5}
\end{eqnarray}
Here $\overline{\rho }_{\mathrm{eq}}=(\rho _{\mathrm{eq},n}+\rho _{\mathrm{eq%
},p})/2$,

\begin{equation}
\mu_{F}={\frac{3}{2}}\,\,\epsilon _{F}\,\rho _{\mathrm{eq}}\frac{s_{R}^{2}}{%
1+F_{1}^{\prime }/3}\left[ 1-{\frac{(1+F_{0}^{\prime })(1+F_{1}^{\prime }/3)%
}{3\,s_{R}^{2}}}\,\right] ,\quad s_{R}=\frac{\omega _{R}}{v_{F}\ q},
\label{B6}
\end{equation}
and $\vec{\chi}$ is the displacement field related to the velocity field $\vec{u}$. 
Namely,
\begin{equation}
\frac{\partial}{\partial t}\vec{\chi}(\vec{r},t)=-\ (1+\kappa)\vec{u}(\vec{r},t),
\quad
\vec{u}(\vec{r},t)=\frac{1}{\rho_{\mathrm{eq}}}\int {\frac{gd\vec{p}}{{(2\pi\hbar)^{3}}}
\frac{\vec{p}}{m}}~\delta f_{\_}(\vec{r},\vec{p}
,t).  \label{B7}
\end{equation}
Using Eqs. (\ref{B2}) - (\ref{B5}), we obtain
\begin{equation}
\vec{n}\cdot \vec{F|}_{S}=\left[ \left( C_{\mathrm{sym}}\,\overline{\rho }_{%
\mathrm{eq}}-\frac{2}{3}\mu _{F}\right) \vec{\nabla}\cdot \vec{\chi}\ +\
2\mu _{F}\frac{\partial }{\partial r}(\vec{n}\cdot \vec{\chi})\right]
_{r=R_{0}}.  \label{B8}
\end{equation}

To evaluate the isovector surface tension force $\vec{F}_{S}$ we will
consider the variation $\delta E_{S,\mathrm{sym}}$ of the surface symmetry
energy caused by the isovector polarization at the nuclear surface \cite%
{mysw74}
\begin{equation}
\delta E_{S,\mathrm{sym}}=\frac{1}{3}\rho _{\mathrm{eq}}r_{0}Q_{\mathrm{sym}%
}\int dS\ \xi ^{2},  \label{B9}
\end{equation}%
where $Q_{\mathrm{sym}}$ is the coefficient related to the volume, $b_{%
\mathrm{sym,vol}}$, and surface, $b_{\mathrm{sym,surf}}$, terms entering to
the symmetry energy $E_{\mathrm{sym}}$ in the mass formula%
\begin{equation}
Q_{\mathrm{sym}}=\frac{9}{8}\frac{b_{\mathrm{sym,vol}}^{2}}{b_{\mathrm{sym,surf}}},
\quad
E_{\mathrm{sym}}=\frac{1}{2}\frac{(N-Z)^{2}}{A}\left( b_{\mathrm{sym,vol}}
-b_{\mathrm{sym,surf}}\ A^{-1/3}\right) .
\label{B10}
\end{equation}
In Eq. (\ref{B9}), the parameter $\xi$ is the dynamic isovector shift of
neutron-proton spheres in units of $r_{0}$
\begin{equation}
\xi=\frac{1}{r_0}\left[R_n(t)-R_p(t)\right]=\frac{\delta R_1(t)}{r_0},  
\label{B11}
\end{equation}
where
\begin{equation}
\delta R_{1}(t)=R_{0}\alpha _{S}(t)Y_{10}(\hat{r}).  \label{B12}
\end{equation}%
The amplitude $\alpha _{S}(t)$ of the isovector shift of the nuclear surface
is connected with the displacement field $\vec{\chi}$. To establish this
connection we note that, for the case of sharp nuclear surface, the
isovector displacement field $\vec{\chi}$ is given by \cite{bomo75}
\begin{equation}
\vec{\chi}(\vec{r},t)=\alpha _{1}(t)\frac{1}{q^{2}}\vec{\nabla}_{\vec{r}}%
\left[ j_{1}(qr)Y_{10}(\hat{r})\right] .  \label{B13}
\end{equation}%
The boundary condition for the normal component of the velocity field reads%
\begin{equation}
\vec{n}\cdot \vec{u}|_{S}=\frac{\partial }{\partial t}\delta R_{1}(t).
\label{B14}
\end{equation}%
Using Eqs. (\ref{B7}), (\ref{B12}) and (\ref{B14})
\begin{equation}
\alpha_{S}(t)=-\ \alpha _{1}(t)\frac{j_{1}^{\prime}(x)}{x(1+\kappa)}, \qquad x=qR_{0}.
\label{B15}
\end{equation}

The variation $\delta E_{S,\mathrm{sym}}$ of the surface energy derives the
surface pressure%
\begin{equation}
\delta P_{S}=\frac{\partial }{\partial \ \delta R_{1}}\frac{\delta E_{S}}
{\delta S}=\frac{8}{3}\frac{\rho_{\mathrm{eq}}}{r_{0}}Q_{\mathrm{sym}}\
\delta R_{1}(t).
\label{B16}
\end{equation}
Taking into account Eqs. (\ref{B12}), (\ref{B15}) and (\ref{B16})
\begin{equation}
\vec{n}\cdot\vec{F}_{S}=-\ \delta P_{S}=\frac{8}{3}\frac{\rho_{\mathrm{eq}}\
j_{1}^{\prime}(x)}{qr_{0}(1+\kappa)}Q_{\mathrm{sym}}\alpha_{1}(t)Y_{10}(\hat{r}).
\label{B17}
\end{equation}
Inserting Eqs. (\ref{B8}) and (\ref{B17}) into Eq. (\ref{B1}) and using Eq.
(\ref{B13}), we obtain the following secular equation
\begin{equation}
\left[-\ \frac{1}{2}C_{\mathrm{sym}}\overline{\rho}_{\mathrm{eq}}
-\frac{2}{3} \mu_{F}+\frac{2}{x^{2}}\mu_{F}\right]j_{1}(x)
+\left[-\ \frac{2}{x}\mu_{F}+\frac{4}{3}\frac{\rho_{\mathrm{eq}}}
{qr_{0}(1+\kappa)}Q_{\mathrm{sym}}\right]j_{1}^{\prime }(x)=0.
\label{B18}
\end{equation}

\end{document}